\documentclass[article]{elsarticle}

\makeatletter
\def\ps@pprintTitle{%
 \let\@oddhead\@empty
 \let\@evenhead\@empty
 \def\@oddfoot{}%
 \let\@evenfoot\@oddfoot}
\makeatother

\usepackage{epsfig,amsmath,amssymb,mathrsfs,color}
\usepackage[active]{srcltx}
\usepackage[matrix,frame,arrow]{xypic} 
\usepackage{amsmath,amssymb,amsfonts}
\usepackage{ifsym}
\usepackage{amsthm}
\usepackage{xypic}
\usepackage{graphicx}% Include figure files
\usepackage{dcolumn}% Align table columns on decimal point
\usepackage{bm}% bold math
\usepackage{lineno,hyperref}
\modulolinenumbers[100]
\usepackage[matrix,frame,arrow]{xy}
\usepackage{amsmath}

\newcommand{\qw}[1][-1]{\ar @{-} [0,#1]}
\newcommand{\gate}[1]{*{\xy *+<.6em>{#1};p\save+LU;+RU **\dir{-}\restore\save+RU;+RD **\dir{-}\restore\save+RD;+LD **\dir{-}\restore\POS+LD;+LU **\dir{-}\endxy} \qw}
\newcommand{\measureD}[1]{*{\xy*+=+<.5em>{\vphantom{\rule{0em}{.1em}#1}}*\cir{r_l};p\save*!R{#1} \restore\save+UC;+UC-<.5em,0em>*!R{\hphantom{#1}}+L **\dir{-} \restore\save+DC;+DC-<.5em,0em>*!R{\hphantom{#1}}+L **\dir{-} \restore\POS+UC-<.5em,0em>*!R{\hphantom{#1}}+L;+DC-<.5em,0em>*!R{\hphantom{#1}}+L **\dir{-} \endxy} \qw}
\newcommand{\multimeasureD}[2]{*+<1em,.9em>{\hphantom{#2}}\save[0,0].[#1,0];p\save !C *{#2},p+LU+<0em,0em>;+RU+<-.8em,0em> **\dir{-}\restore\save +LD;+LU **\dir{-}\restore\save +LD;+RD-<.8em,0em> **\dir{-} \restore\save +RD+<0em,.8em>;+RU-<0em,.8em> **\dir{-} \restore \POS !UR*!UR{\cir<.9em>{r_d}};!DR*!DR{\cir<.9em>{d_l}}\restore \qw}
\newcommand{\multigate}[2]{*+<1em,.9em>{\hphantom{#2}} \qw \POS[0,0].[#1,0];p !C *{#2},p \save+LU;+RU **\dir{-}\restore\save+RU;+RD **\dir{-}\restore\save+RD;+LD **\dir{-}\restore\save+LD;+LU **\dir{-}\restore}
\newcommand{\ghost}[1]{*+<1em,.9em>{\hphantom{#1}} \qw}
\newcommand{\Qcircuit}[1][0em]{\xymatrix @*=<#1>} % @*[o]
\newcommand{\pureghost}[1]{*+<1em,.9em>{\hphantom{#1}}}
\newcommand{\multiprepareC}[2]{*+<1em,.9em>{\hphantom{#2}}\save[0,0].[#1,0];p\save !C
  *{#2},p+RU+<0em,0em>;+LU+<+.8em,0em> **\dir{-}\restore\save +RD;+RU **\dir{-}\restore\save
  +RD;+LD+<.8em,0em> **\dir{-} \restore\save +LD+<0em,.8em>;+LU-<0em,.8em> **\dir{-} \restore \POS
  !UL*!UL{\cir<.9em>{u_r}};!DL*!DL{\cir<.9em>{l_u}}\restore}
\newcommand{\prepareC}[1]{*{\xy*+=+<.5em>{\vphantom{#1\rule{0em}{.1em}}}*\cir{l^r};p\save*!L{#1} \restore\save+UC;+UC+<.5em,0em>*!L{\hphantom{#1}}+R **\dir{-} \restore\save+DC;+DC+<.5em,0em>*!L{\hphantom{#1}}+R **\dir{-} \restore\POS+UC+<.5em,0em>*!L{\hphantom{#1}}+R;+DC+<.5em,0em>*!L{\hphantom{#1}}+R **\dir{-} \endxy}}
\newcommand{\poloFantasmaCn}[1]{{{}^{#1}_{\phantom{#1}}}}

\def\<{\langle}\def\>{\rangle} 
 
\def\transf#1{\mathcal{#1}}
\def\tE{\transf{E}}\def\tF{\transf{F}}\def\tG{\transf{G}}
\def\tH{\transf{H}}\def\tI{\transf{I}}
\def\tA{\transf{A}}\def\tB{\transf{B}}\def\tC{\transf{C}}\def\tD{\transf{D}}  \def\tS{\transf{S}}
\def\tU{\transf{U}}\def\tT{\transf{T}}
\def\sys#1{\mathrm{#1}}
\def\rX{\sys{X}}
\def\rM{\sys{M}}
\def\rN{\sys{N}}
\def\rA{\sys{A}}\def\rB{\sys{B}}\def\rC{\sys{C}}
\def\rD{\sys{D}}\def\rY{\sys{Y}}   \def\rI{\sys{I}}
\def\rE{\sys{E}}
\def\St{\mathsf{St}}\def\Eff{\mathsf{Eff}}
\def\Transf{\mathsf{Transf}}
\def\Outcomes{\mathsf{Outcomes}}
\def\Test{\mathsf{Tests}}
\def\Det{\mathsf{Det}}
\def\Purs{\mathsf{Pur}}
\def\Pur{\mathsf{Pur}}
\def\Rev{\mathsf{Rev}}
\def\Finite{\mathsf{Finite}}
\def\Op{\mathsf{Op}}
\def\Prob{\mathsf{Prob}}
\def\Sys{\mathsf{Sys}}
\def\Str{\St_{\mathbb R}}\def\Effr{\Eff_{\mathbb
    R}}
\def\Stp{\St_+}\def\Effp{\Eff_+}
\def\Stn{\St_1}
\def\over{\overline}
\def\R{{\mathbb R}}
\def\N{{\mathbb N}}
\def\Tr{\operatorname{Tr}}
\def\Span{\mathsf{Span}}
\def\bs#1{\boldsymbol{#1}}

\newtheorem{postulate}{Postulate}
\newtheorem{prop}{Proposition}
\newtheorem{theorem}{Theorem}
\newtheorem{cor}{Corollary}
\newtheorem{lemma}{Lemma}
\newtheorem{definition}{Definition}

\def\rA{{\rm A}}\def\rB{{\rm B}}\def\rC{{\rm C}}\def\rD{{\rm D}}
\def\rE{{\rm E}}   \def\rI{{\rm
    I}} \def\rR{{\rm R}} \def\rS{{\rm S}} \def\rX{{\rm
    X}} \def\rY{{\rm Y}}\def\rZ{{\rm Z}}

\begin{document}

\begin{frontmatter}

\title{Quantum from principles}

\author{Giulio Chiribella\footnote{Corresponding author: giulio@cs.hku.hk}}
\address{%Department of Computer Science, The University of Hong Kong, Hong Kong, P. R. China
Center for Quantum Information, Institute for Interdisciplinary Information Sciences, Tsinghua University, Beijing, 100084}

\author{Giacomo Mauro D'Ariano}
\address{{\em QUIT Group}, Dipartimento di Fisica, Universit\`a di Pavia, via Bassi 6, 27100 Pavia, Italy}
\address{INFN sezione di Pavia, via Bassi 6, 27100 Pavia, Italy}
%\homepage{http://www.qubit.it}
%\email{dariano@unipv.it}

\author{Paolo Perinotti}
\address{{\em QUIT Group}, Dipartimento di Fisica, Universit\`a di Pavia, via Bassi 6, 27100 Pavia, Italy}
\address{INFN sezione di Pavia, via Bassi 6, 27100 Pavia, Italy}
%\homepage{http://www.qubit.it}
%\email{perinotti@unipv.it}




\begin{abstract}
 Quantum theory was discovered  in an adventurous way,  under the urge to  solve puzzles---like the spectrum of the  blackbody radiation---that haunted  the physics community at
  the beginning of the 20th century.   It soon became clear, though,  that quantum theory was not just a theory of specific physical systems, but rather a new language of universal applicability.  Can this language be reconstructed from first principles?   Can we arrive at it  from logical reasoning, instead of \emph{ad hoc} guesswork?  
  A positive answer was provided in  Refs.  \cite{puri,deri}, where we put forward six principles that identify  quantum theory uniquely in a broad class of theories. 
 We first  defined a class of  ``theories of information",  constructed  as  extensions of probability theory in which   events can be connected into networks. In this framework, we formulated  the six principles  as  rules governing  the control and the accessibility of information.    Directly from these rules, we reconstructed  a number of quantum information features, and eventually, the whole  Hilbert space framework.    In short, our principles characterize  quantum theory as the theory of information that allows for  maximal control of randomness. 
 \end{abstract}

\end{frontmatter}

\linenumbers

\section{Introduction}

Quantum foundations is an old field---as old as quantum mechanics itself.  Among the early works
stand out the seminal papers by Einstein, Podolski, and Rosen \cite{epr} and Schr\"odinger \cite{schrodinger1935discussion},
who addressed quantum entanglement for the first time, exploring quantum mechanics   
%``from the inside'' 
within the Hilbert space formulation.  Almost at the same time, Birkhoff and von Neumann
\cite{birkvon} looked at the theory %``from the outside'', 
in a wider framework allowing for
alternative theories.  From that angle, it was natural to ask what is special about quantum mechanics
and why Nature obeys its peculiar laws.  The take of Birkhoff and von Neumann was that quantum
theory should be regarded as a new form of logic, whose laws could be derived from physically motivated axioms.  This  programme gave rise to the tradition of quantum logic
\cite{mackey,ludwig,piron,jauchpiron,beltrametticassinelli}, whose ramifications are still object of
active research \cite{bobquantumlogic}.

Researchers in quantum logic managed to derive a significant part of the quantum framework from
logical axioms. However, there is a general consensus that the axioms put forward in this context
are not as insightful as one would have hoped. For both experts and non-experts, it is hard to
figure out what is the moral of the quantum-logic axiomatizations. What is special about quantum
theory after all?  Why should quantum theory be preferred to alternative theories?  Not many answers
can be found in the popular accounts of quantum logic (see e.g. the Wikipedia entry \cite{wikipedia}) and even understanding what the
axioms are requires delving into a highly specialized literature.
 
The ambition to find a more insightful axiomatization  reemerged with the rise of
quantum information.  The new field  showed that the mathematical axioms of quantum theory imply striking operational consequences, such as quantum key distribution \cite{bb84,e91},
quantum algorithms \cite{shor,grover}, no-cloning \cite{wootzu,dieks1982communication}, quantum teleportation
\cite{telep} and dense coding \cite{densec}.  A natural question is: Can we reverse the implication
and \emph{derive} the mathematics of quantum theory from some of its operational consequences?  This
question is at the core of a research programme launched by Fuchs \cite{littlemore} and Brassard
\cite{cryptofound}, which can be synthesized by the motto \emph{``quantum foundations in the light of
  quantum information"}   \cite{fuchs2001quantum} \footnote{This was also the title of one influential conference, held in May 2000 at the Universit\'e de Montr\'eal \cite{fuchs2011coming}, which kickstarted the new wave of quantum axiomatizations. }.  The ultimate goal of the programme is to reconstruct the whole structure of quantum theory from few simple principles of information-theoretic nature. 

One may wonder why quantum information theorists should be more successful than their predecessors in the axiomatic endeavour.  A good reason is the following: 
In the pre-quantum information era, quantum theory was viewed like an impoverished version of
classical theory, lacking the ability to make deterministic predictions about the outcomes of
experiments.  Clearly, this perspective offered no vantage point for explaining why the world should
be quantum.  Contrarily, quantum information provided plenty of positive reasons for preferring quantum
theory to its classical counterpart.   Turning some of these reasons
into axioms then appeared as a promising route towards a compelling axiomatization.  Pioneering works
along this route are those by Hardy \cite{hardy01} and D'Ariano \cite{maurofirst,maurobook}.  More
recently, the programme flourished, leading to an explosion  of new axiomatizations
%leiferbarnumshortwilce,barrett,puri,
\cite{deri,hardy11,masanes11,dakic11, goyal10, masanes12,wilce2012conjugates,barnum2014higher}.
  %, each one telling a different story about quantum theory.   

Here we review the axiomatization of Refs. \cite{deri}.   In this work, quantum theory is derived from six  principles, formulated in a general framework of theories of information. 
The first five principles---Causality,  Purity of Composition, Local Discriminability,  Perfect State Discrimination, and Ideal Compression---express ordinary properties  that are shared by  quantum and classical  information theory: such principles define what we call a \emph{standard} theory of information.   Among all standard theories of information, the sixth principle---Purification---identifies quantum theory uniquely.    Purification states that every random preparation can be simulated via non-random preparation procedure, in which the  system is manipulated together with an environment.  An agent that has access to both the system and the environment would then have maximal  control of the preparation---\emph{maximal} in the sense that no other agent could conceivably have higher control.     
%At the level of processes, Purification implies that every deterministic process can be simulated through a reversible interaction between the system and its environment, with the environment in a non-random state, completely under the agent's control.  
The moral of our work is that {\em Quantum   Theory is the theory that allows maximal control of randomness},  giving us---at least in principle---the power to control all possible preparations and all possible  dynamics.

The chapter is structured as follows:  in section  \ref{sec:OPTs} we provide an introduction to the framework of \emph{operational-probabilistic theories}---general theories of information arising from the combination of the circuit framework with probability theory.   Then, section \ref{sec:background} presents the background to the reconstruction, discussing the main standing assumptions---finite-dimensionality, non-determinism, and closure under limits---and introducing  a few  basic operational tasks:  signalling, collecting side information, doing state tomography, distinguishing states, compressing information, and simulating preparations.   
The principles are then analyzed in section \ref{sec:ax}.    Section \ref{sec:reconstruction} provides a guided tour through the main results in our reconstruction, showing how the main features of quantum theory can be derived directly from the principles.  Finally, the conclusions are drawn in section \ref{sec:conclusions}.

\section{Operational-probabilistic theories}\label{sec:OPTs}

In order to reconstruct quantum theory and the features of quantum information, one needs a framework capable to describe a variety of alternative theories.   Different frameworks have been proposed for this scope, under the broad name of \emph{general probabilistic  theories}   \cite{hardy01,barrett07,maurofirst,maurobook,nobroad,barnum08,puri,deri,hardy11,hardy2013,barnum11}.  Our reconstruction is based on  a specific variant of general probabilistic theories, which we call  \emph{operational-probabilistic theories (OPTs)}  \cite{puri,deri}. 
  OPTs are an extension of probability theory, in which events can be connected into circuits.  Technically, OPTs arise from the combination of the categorical framework  of Abramsky and Coecke     \cite{abramsky2004,abramsky2008,coecke2010universe}  with the toolbox of elementary probability theory.    We regard such a combination as the natural mathematical object describing a ``general theory of information". In the following we present a concise summary of the OPT framework.

\subsection{Operational structure}

\subsubsection{Systems}

Systems are labels,  which determine how different events can be connected to one another.   We denote systems by  capital letters, such as $\rA,\rB, \rC$, and so on. 
The letter $\rI$ will be reserved for the \emph{trivial system}, representing  ``nothing" \footnote{More precisely, ``nothing that the theory cares to describe".  }.  The set of all systems under consideration will be denoted by $\Sys$.  
  
%  In quantum theory, systems are associated to complex Hilbert spaces and the trivial system is associated to the one-dimensional Hilbert space  $\mathbb C$.   

Every two systems $\rA$ and $\rB$ can be considered together as a composite system, denoted by $\rA\otimes \rB$.  The composition of systems is associative, namely
 \begin{align}\label{ass1} \rA \otimes (\rB\otimes \rC)   =   (\rA\otimes \rB)  \otimes \rC    \qquad \forall \rA,\rB  , \rC  
 \end{align} 
and has the trivial system as identity element, namely
\begin{align}\label{unit}  \rA\otimes \rI   =    \rI\otimes \rA  =  \rA  \qquad \forall \rA\,.  
\end{align}
The second condition means that considering  system $\rA$ together with ``nothing"  is the same as considering system $\rA$ alone.

\subsubsection{Events}  An event of type $\rA\to \rB$ represents the occurrence of a  transformation that  converts the input system $\rA$ into the output system $\rB$.     An event  $\tE$  of type $\rA\to \rB$   will be represented graphically as
\begin{equation}
\Qcircuit @C=1em @R=.7em @! R {
& \poloFantasmaCn{\rA} \qw &\gate{  \tE}&\poloFantasmaCn{\rB} \qw &\qw}\nonumber \, .
\end{equation}  
The set of all events  of type $\rA\to \rB$ will be denoted by $\Transf(\rA\to \rB)$, identifying events with the corresponding transformations.

When the input and output systems are composite systems, we draw boxes with multiple wires. For example, the box  
\[  
\begin{aligned}
\Qcircuit @C=1em @R=.7em @! R {
& \poloFantasmaCn{\rA} \qw &\multigate{1}{  \tE}&\poloFantasmaCn{\rB} \qw &\qw\\
& \poloFantasmaCn{\rC} \qw &\ghost{  \tE}&\poloFantasmaCn{\rD} \qw &\qw}
 \end{aligned} 
 \   :=   ~  
 \begin{aligned}  \Qcircuit @C=1em @R=.7em @! R {
&\qw & \poloFantasmaCn{\rA\otimes \rC \quad} \qw& \gate{  \tE}&\poloFantasmaCn{ \quad \rB\otimes \rD} \qw &\qw&\qw}\end{aligned}  \]
represents an event of type $(\rA\otimes  \rC)\to (\rB\otimes \rD)$.  

 Some types of  events are particularly important and deserve a name of their own.   An event of type $\rI \to \rA$ is a  {\em preparation-event} (or simply, a \emph{preparation}), that is, an event that makes system $\rA$ available to further processing.    
 An event of type $\rA\to \rI$  is an {\em observation-event} (or simply, an \emph{observation}), after which system $\rA$  is no longer available.   Preparation- and observation-events will be represented as 
\[
  \begin{aligned}
    \Qcircuit @C=1em @R=.7em @! R {
      &\prepareC{\rho}&\poloFantasmaCn{\rA}\qw  &\qw }\ :=  \quad \Qcircuit
    @C=1em @R=.7em @! R {
      &\qw \poloFantasmaCn{\rI}&\gate{\rho }&\poloFantasmaCn{\rA}\qw  &\qw }
      \end{aligned}
  \] 
  and 
  \[
  \begin{aligned} \Qcircuit @C=1em @R=.7em @! R {
      &\qw \poloFantasmaCn{\rA}&\measureD{m}}\ :=~ \Qcircuit
    @C=1em @R=.7em @! R {
      &\qw \poloFantasmaCn{\rA}&\gate{m}&\poloFantasmaCn{\rI}\qw  &\qw}
  \end{aligned} ~,
  \]
respectively.    We will often use the Dirac-like notation $(a|$ and  $|\rho)$  the observation $a$ and  the preparation $\rho$, respectively.  

Events of type $\rI\to \rI$  will be called \emph{scalars}  \cite{abramsky2004}.  Scalars  will be represented ``out of the box", as 
 \[
  s  :=~ 
  \begin{aligned}\Qcircuit
    @C=1em @R=.7em @! R {
      &\qw \poloFantasmaCn{\rI}&\gate{s}&\poloFantasmaCn{\rI}\qw  &\qw}
  \end{aligned} ~.
  \]
Later, scalars will be associated to probabilities. For the moment, however,  they are just a special type of events.

\subsubsection{Composition of events} 
 Events can be connected into  networks through the following operations
 \begin{enumerate}
\item \emph{Sequential composition:} an event of type $\rA\to \rB$ can be connected to  an event  of type $\rB\to \rC$, yielding an event of type $\rA \to \rC$.  
\item \emph{Parallel composition:}  an event of type $\rA\to \rA'$ can be composed with an event  of type $\rB\to \rB'$, yielding an event of type $(\rA \otimes \rB)\to (\rA'\otimes \rB')$.
\end{enumerate}  

 Intuitively, the sequential composition represents two events happening at ``subsequent time steps"  \footnote{{\em Per se}, the mathematical formalism does not force us to interpret the order of sequential composition as an  order in time. Nevertheless, composition in time is the reference situation that we will have  in mind when phrasing our axioms.  }.  The sequential composition of two events $\tE$ and $\tF$ of matching types is  denoted by $ \tF\circ\tE$ and is represented graphically as 
\[
\begin{aligned}\Qcircuit @C=1em @R=.7em @! R {
& \poloFantasmaCn{\rA} \qw &\gate{  \tE}&\poloFantasmaCn{\rB} \qw &\gate{\tF}  &\poloFantasmaCn{\rC} \qw  &\qw}
\end{aligned}  
\   :=    ~
\begin{aligned}
\Qcircuit @C=1em @R=.7em @! R {
& \poloFantasmaCn{\rA} \qw &\gate{ \tF\circ  \tE}&\poloFantasmaCn{\rC} \qw &\qw} 
\end{aligned} ~ \, .
\]
%(note that the order of sequential composition is from left to right  in the picture $\Qcircuit @C=1em @R=.7em @! R {
%& \poloFantasmaCn{\rA} \qw &\gate{  \tE}&\poloFantasmaCn{\rB} \qw &\gate{\tF}  &\poloFantasmaCn{\rC} \qw  &\qw}$, while in the formula).    
This graphical  notation is justified  by  the requirement that sequential composition be associative, namely   
\begin{align}\label{ass2} 
\tG  \circ   (\tF  \circ \tE)     = ( \tG  \circ   \tF ) \circ \tE   \, ,       
\end{align}
for arbitrary events $\tE,\tF$ and $\tG$ of matching types.  In addition to associativity, sequential composition is required to have an identity element for every system.    The \emph{identity on system    $\rA$}, denoted by $\tI_\rA$, is the special event  of type $\rA\to \rA$ identified by the conditions  
\begin{align}\label{id1} \begin{aligned}\Qcircuit @C=1em @R=.7em @! R {
& \poloFantasmaCn{\rA} \qw &\gate{  \tI_\rA}&\poloFantasmaCn{\rA} \qw &\gate{\tE}  &\poloFantasmaCn{\rB} \qw  &\qw}
\end{aligned}  
    =    \begin{aligned}
\Qcircuit @C=1em @R=.7em @! R {
& \poloFantasmaCn{\rA} \qw &\gate{  \tE}&\poloFantasmaCn{\rB} \qw &\qw}       \end{aligned}\end{align}   
and
\begin{align}\label{id2}\begin{aligned}\Qcircuit @C=1em @R=.7em @! R {
& \poloFantasmaCn{\rB} \qw &\gate{  \tF}&\poloFantasmaCn{\rA} \qw &\gate{\tI_\rA}  &\poloFantasmaCn{\rA} \qw  &\qw}
\end{aligned}  
    =    \begin{aligned}
\Qcircuit @C=1em @R=.7em @! R {
& \poloFantasmaCn{\rB} \qw &\gate{  \tF}&\poloFantasmaCn{\rA} \qw &\qw}   
\end{aligned}~ ,\end{align}   
required to be valid  for arbitrary systems  $\rA,\rB$ and  arbitrary events $\tE$ and $\tF$ of types $\rA\to \rB$  and $\rB\to \rA$, respectively.   The intuitive content of the above equations is that $\tI_\rA$   represents the process that ``does nothing on the system".
Consistently,   we use the graphical  notation 
\[   \begin{aligned}
\Qcircuit @C=1em @R=.7em @! R {
& \qw &\qw \poloFantasmaCn{\rA}  &\qw&\qw}   
\end{aligned}    \ :=  ~       \begin{aligned}
\Qcircuit @C=1em @R=.7em @! R {
& \poloFantasmaCn{\rA} \qw &\gate{  \tI_\rA}&\poloFantasmaCn{\rA} \qw &\qw}   ~. 
\end{aligned}  \]
Mathematically,   conditions (\ref{ass2}), (\ref{id1}), and (\ref{id2}) impose that  
  the  events  form a \emph{category}  \cite{coecke2011categories,mac1978categories}, in which the systems are the objects and the events are the arrows.          For the sequential composition of a preparation and an observation we will often use the Dirac-like notation,  
\begin{align}
(a|\rho)   :   =
\begin{aligned}
\Qcircuit @C=1em @R=.7em @! R {
&   \prepareC{\rho}   &  \poloFantasmaCn{\rA} \qw & \measureD{a}  }
\end{aligned} ~ .
\end{align} 
%Intuitively, sequential composition represents events occurring ``one after the other".  

Let us  consider  parallel composition.
 The parallel composition of two events $\tE$ and $\tF$  is  denoted as $\tE  \otimes \tF$   and is  represented graphically as 
\[
\begin{aligned}\Qcircuit @C=1em @R=.7em @! R {
& \poloFantasmaCn{\rA} \qw   &\gate{\tE}   &\poloFantasmaCn{\rA'} \qw &\qw \\
&  \poloFantasmaCn{\rB}  \qw &\gate{\tF}  &\poloFantasmaCn{\rB'} \qw  &\qw}
\end{aligned}  
\   :=    ~
\begin{aligned}
\Qcircuit @C=1em @R=.7em @! R {
& \poloFantasmaCn{\rA} \qw &\multigate{1}{ \tE\otimes  \tF}&\poloFantasmaCn{\rA'} \qw &\qw\\
& \poloFantasmaCn{\rB} \qw &\ghost{ \tE\otimes  \tF}&\poloFantasmaCn{\rB'} \qw &\qw} 
\end{aligned}  ~.
\]
The graphical notation is justified by the requirement of the following   condition  
\begin{align}\label{consistency}
 ( \tE\otimes \tF  )  \circ  (  \tG  \otimes \tH)   =    ( \tE  \circ \tG  )   \otimes  (  \tF\circ \tH ) \, , 
\end{align}
where $\tE,\tF,\tG,$ and $\tH$ are arbitrary events of matching types. Such  condition is necessary for the graphical notation to make sense, since in graphical  notation the two sides of Eq. (\ref{consistency}) look exactly the same.    
\iffalse
Note that the condition  (\ref{consistency}), combined with Eqs. (\ref{id1}) and (\ref{id2}),  implies the relation  \[(\tI_{\rA'}  \otimes  \tF)  \circ(\tE\otimes  \tI_\rB)   =  (  \tE  \otimes \tI_{\rB'})  \circ  ( \tI_\rA\otimes \tF )  \, ,\] graphically expressed as  
\[
\begin{aligned}\Qcircuit @C=1em @R=.7em @! R {
& \poloFantasmaCn{\rA} \qw   &\gate{\tE}   & \qw &\qw  &\qw   \poloFantasmaCn{\rA'} &\qw  \\
&   \qw  \poloFantasmaCn{\rB} &\qw   &   \qw &\gate{\tF}  &\poloFantasmaCn{\rB'} \qw  &\qw}
\end{aligned}  
  \  =  ~  
  \begin{aligned}
  \Qcircuit @C=1em @R=.7em @! R {
&   \qw  \poloFantasmaCn{\rA}    &\qw   & \qw &\gate{\tE}  &\poloFantasmaCn{\rA'} \qw  &\qw\\
& \poloFantasmaCn{\rB} \qw   &\gate{\tF}   &\qw &\qw  &\qw  \poloFantasmaCn{\rB'}  &\qw}
\end{aligned}   \, .
\]
In words: when two events happen in parallel, it does not matter which of the two happens first.   

\fi 
%Note that the validity of the equation is suggested already by the graphical language, where composing an event with the identity is equivalent to ``stretching a wire".
In addition to Eq. (\ref{consistency}),  parallel composition is required to satisfy the condition 
\begin{align}\label{identityprod}
 \tI_{\rA\otimes \rB}  =   \tI_\rA  \otimes \tI_\rB \, .      
\end{align}
Mathematically, the presence of parallel composition turns the category of events into a strict monoidal category, whose key properties are  summarized by Eqs. (\ref{ass1}), (\ref{unit}), (\ref{consistency}), and (\ref{identityprod}). We denote such category by $\Transf$.

\subsubsection{Reversible events}  An event  $\tE$ of type $\rA\to \rB$ is \emph{reversible} iff there exists 
 another event $\tF$, of type $\rB\to \rA$, such that  
 \begin{align}\label{rev1}\begin{aligned}\Qcircuit @C=1em @R=.7em @! R {
& \poloFantasmaCn{\rA} \qw &\gate{  \tE}&\poloFantasmaCn{\rB} \qw &\gate{\tF}  &\poloFantasmaCn{\rA} \qw  &\qw}
\end{aligned}  
    =    \begin{aligned}
\Qcircuit @C=1em @R=.7em @! R {
&\qw &  \poloFantasmaCn{\rA} \qw &  \qw&\qw }   
\end{aligned}~ ,\end{align} 
and   
 \begin{align}\label{rev2}\begin{aligned}\Qcircuit @C=1em @R=.7em @! R {
& \poloFantasmaCn{\rB} \qw &\gate{  \tF}&\poloFantasmaCn{\rA} \qw &\gate{\tE}  &\poloFantasmaCn{\rB} \qw  &\qw}
\end{aligned}  
    =    \begin{aligned}
\Qcircuit @C=1em @R=.7em @! R {
&\qw &  \poloFantasmaCn{\rB} \qw &  \qw&\qw }   
\end{aligned}~ .\end{align} 
 When this is the case, we write $\tF  =  \tE^{-1}$ and we say that  systems $\rA$ and $\rB$ are \emph{operationally equivalent}  (or simply {\em equivalent}).  

We denote by $\Rev\Transf(\rA\to \rB)$ the set of reversible events of type $\rA\to \rB$.    Such set (which may be empty)   depends on the specific theory.  
In general, we require the existence of a reversible event that swaps  pairs of systems.  Given two systems $\rA$ and $\rB$, the \emph{swap of $\rA$ with $\rB$}---denoted by $\tS_{\rA,\rB}$---is a reversible    event of type $(\rA\otimes \rB)  \to (\rB\otimes \rA)$ satisfying the  condition  
\begin{align}\label{swap1}
%\tS_{\rA,\rB}^{-1}   =  \tS_{\rB,\rA} \, .
\begin{aligned}
\Qcircuit @C=1em @R=.7em @! R {
& \poloFantasmaCn{\rA} \qw &\multigate{1}{  \tS_{\rA,\rB}}&\poloFantasmaCn{\rB} \qw &  \gate{  \tF}   &  \poloFantasmaCn{\rB'} \qw   &   \multigate{1}{  \tS_{\rB',\rA'}}&\poloFantasmaCn{\rA'} \qw &\qw  \\
& \poloFantasmaCn{\rB} \qw &\ghost{  \tS_{\rA,\rB}  }&\poloFantasmaCn{\rA} \qw &  \gate{\tE}  &  \poloFantasmaCn{\rA'} \qw   &   \ghost{  \tS_{\rB',\rA'}}&\poloFantasmaCn{\rB'} \qw &\qw }
\end{aligned}  \  =  ~  
\begin{aligned}\Qcircuit @C=1em @R=.7em @! R {
& \poloFantasmaCn{\rA} \qw   &\gate{\tE}   &\poloFantasmaCn{\rA'} \qw &\qw \\
&  \poloFantasmaCn{\rB}  \qw &\gate{\tF}  &\poloFantasmaCn{\rB'} \qw  &\qw}
\end{aligned}   \, ,
 \end{align}
for arbitrary systems $\rA,\rB,\rA',\rB'$ and arbitrary events $\tE,\tF$, as well as the conditions  
\begin{align}\label{swap2}
%\tS_{\rA,\rB}^{-1}   =  \tS_{\rB,\rA} \, .
\begin{aligned}
\Qcircuit @C=1em @R=.7em @! R {
& \poloFantasmaCn{\rA} \qw &\multigate{1}{  \tS_{\rA,\rB}}&\poloFantasmaCn{\rB} \qw & \multigate{1}{  \tS_{\rB,\rA}}&\poloFantasmaCn{\rA} \qw &\qw  \\
& \poloFantasmaCn{\rB} \qw &\ghost{  \tS_{\rA,\rB}  }&\poloFantasmaCn{\rA} \qw &\ghost{  \tS_{\rB,\rA}}&\poloFantasmaCn{\rB} \qw &\qw }
\end{aligned}  \  =  ~  
\begin{aligned}
\Qcircuit @C=1em @R=.7em @! R {
&\qw &  \poloFantasmaCn{\rA} \qw &  \qw&\qw \\
&&&&\\
&\qw &  \poloFantasmaCn{\rB} \qw &  \qw&\qw }   
\end{aligned}
\end{align}
and 
\begin{align}\label{swap3}
\begin{aligned}
\Qcircuit @C=1em @R=.7em @! R {
& \poloFantasmaCn{\rA} \qw &\multigate{2}{  \tS_{\rA,\rB\otimes \rC}}&\poloFantasmaCn{\rB} \qw &\qw\\
& \poloFantasmaCn{\rB} \qw &\ghost{  \tS_{\rA,\rB\otimes \rC}}&\poloFantasmaCn{\rC} \qw &\qw\\
& \poloFantasmaCn{\rC} \qw &\ghost{  \tS_{\rA,\rB\otimes \rC}  }&\poloFantasmaCn{\rA} \qw &\qw}
 \end{aligned} 
 \  =  ~
\begin{aligned}
\Qcircuit @C=1em @R=.7em @! R {
& \poloFantasmaCn{\rA} \qw &\multigate{1}{  \tS_{\rA,\rB}}&\poloFantasmaCn{\rB} \qw &\qw&\qw &\qw \\
& \poloFantasmaCn{\rB} \qw &\ghost{  \tS_{\rA,\rB}  }&\poloFantasmaCn{\rA} \qw &   \multigate{1}{  \tS_{\rA,\rC}}  &    \poloFantasmaCn{\rC} \qw &\qw      \\
&  \poloFantasmaCn{\rC} \qw  &  \qw &\qw & \ghost{  \tS_{\rA,\rC}  }& \poloFantasmaCn{\rA} \qw &\qw    }
 \end{aligned}  \, ,
\end{align}
 The presence of the swap, with the related equations (\ref{swap1}), (\ref{swap2}), and (\ref{swap3}),   turns the  strict monoidal category into a  strict  \emph{symmetric} monoidal category \cite{coecke2010quantum,selinger2011survey} (strict SMC, for short).

\subsubsection{Tests}  

A test represents a process, which can generally be non-deterministic---i.~e.~it can result in  multiple alternative events.  Specifically, a test
  of type $\rA\to \rB$ is  collection of  events of type $\rA\to \rB$, labelled by outcomes in a suitable outcome set $\rX$. The test  $\boldsymbol{\tE}  :  =\{\tE_x\}_{x\in\rX}$  is represented graphically as
\begin{equation}
\begin{aligned}
\Qcircuit @C=1em @R=.7em @! R {
& \poloFantasmaCn{\rA} \qw &\gate{  \boldsymbol{\tE}  }&\poloFantasmaCn{\rB} \qw &\qw}
\end{aligned}= 
\begin{aligned}
\Qcircuit @C=1em @R=.7em @! R {
& \poloFantasmaCn{\rA} \qw &\gate{  \{\tE_x\}_{x\in\rX}}&\poloFantasmaCn{\rB} \qw &\qw}
\end{aligned}\nonumber ~ .
\end{equation}   
When two events/transformations belong to the same test, we say that they are \emph{coexisting}.

The set of tests of type $\rA\to \rB$ with outcomes in $\rX$ will be denoted by   $\Test (\rA\to \rB,\rX)$.     We will restrict our attention to tests with a \emph{finite} outcome set.

Tests with $|\rX|  = 1$ are  called \emph{deterministic}, because  only one event can take place.  We will often identify a deterministic test $\{  \tE_{x_0} \}$ with the corresponding event  $\tE_{x_0}$, saying that $\tE_{x_0}$  is a \emph{deterministic event}  (or a \emph{deterministic transformation}).       The deterministic transformations form a strict    symmetric monoidal subcategory of $\Transf$, denoted by $\Det\Transf$.

Some types of tests are particularly important and deserve a name of their own.   A test of type $\rI \to \rA$ is a  {\em preparation-test}  (or an \emph{ensemble}), which  prepares system $\rA$ in a non-deterministic way, with the possible preparations labelled by different outcomes.    
 A test of type $\rA\to \rI$  is an {\em observation-test}, corresponding to a demolition measurement that absorbs  system $\rA$ while producing an outcome.   
\subsubsection{Composition of tests}
 
Not all collections of events of  are ``tests". Whether or not a specific collection  is a test is determined by the theory, compatibly with two basic requirements: 
\begin{enumerate}
\item the set of tests must be closed under sequential and parallel composition
\item the set of tests must contain  deterministic tests corresponding to reversible events.
\end{enumerate} 

Let us discuss these requirements in  more  detail:  

\begin{enumerate}
\item   The sequential composition of two tests $  \boldsymbol{\tE}  = \left\{ \tE_x\right\}_{x\in\rX}$ and  $\boldsymbol{\tF}  =\left\{\tF_y\right\}_{y\in\rY}$ of matching  types is defined as  
\[ \boldsymbol{\tF}  \circ  \boldsymbol{\tE}   :   =   \left\{ \tF_y \circ \tE_x \right\}_{(x,y)  \in  \rX\times \rY} \, . \]
The test  $  \boldsymbol{\tF}  \circ  \boldsymbol{\tE}   $ represents a cascade of two (generally non-deterministic) processes, wherein each process can result  in a number of alternative events.     
Similarly, the parallel composition of two tests is defined as  
\[    \boldsymbol{\tE}  \otimes  \boldsymbol{\tF}     :   =   \left\{  \tE_x  \otimes \tF_y  \right\}_{(x,y)  \in  \rX\times \rY}  \] 
and represents two  non-deterministic processes occurring in parallel.  The composition of tests induces a composition of their outcome spaces via the Cartesian product.  As a consequence, the set of all outcome spaces must be closed under this operation.  We will denote such a set by $\Outcomes$.

\item       If $\tU$ is a reversible event of type $\rA\to \rB$, we require that there exists a deterministic test $\boldsymbol{\tU}  : = \{\tU\}$.  In particular, there must be a deterministic test $\boldsymbol{\tI}_\rA  :=  \{\tI_\rA\}$ corresponding to the identity on system $\rA$ and a deterministic test  $\boldsymbol{\tS}_{\rA,\rB}  := \{\tS_{\rA,\rB} \}$ corresponding to the swap of systems $\rA$ and $\rB$.   
\end{enumerate}

Note that  all the basic equations valid for events can be lifted to tests:  for example,   the identity test acts as identity element with respect to the composition, that is, one has 
\begin{align}\label{idd1} \begin{aligned}\Qcircuit @C=1em @R=.7em @! R {
& \poloFantasmaCn{\rA} \qw &\gate{   \boldsymbol{ \tI}_\rA}&\poloFantasmaCn{\rA} \qw &\gate{\boldsymbol{\tE}}  &\poloFantasmaCn{\rB} \qw  &\qw}
\end{aligned}  
    =    \begin{aligned}
\Qcircuit @C=1em @R=.7em @! R {
& \poloFantasmaCn{\rA} \qw &\gate{ \boldsymbol{\tE} }&\poloFantasmaCn{\rB} \qw &\qw}       \end{aligned}\end{align}   
and
\begin{align}\label{idd2}\begin{aligned}\Qcircuit @C=1em @R=.7em @! R {
& \poloFantasmaCn{\rB} \qw &\gate{ \boldsymbol{ \tF}}&\poloFantasmaCn{\rA} \qw &\gate{\boldsymbol{\tI}_\rA}  &\poloFantasmaCn{\rA} \qw  &\qw}
\end{aligned}  
    =    \begin{aligned}
\Qcircuit @C=1em @R=.7em @! R {
& \poloFantasmaCn{\rB} \qw &\gate{ \boldsymbol{ \tF}}&\poloFantasmaCn{\rA} \qw &\qw}   
\end{aligned}~ ,\end{align}   
for arbitrary systems  $\rA,\rB$ and  for arbitrary tests $\boldsymbol{\tE}$ and $\boldsymbol{\tF}$ of types $\rA\to \rB$  and $\rB\to \rA$,  respectively.     
 Since events form a strict SMC, also the  tests   form a strict SMC, which we denote by $\Test$.

\subsubsection{Summary about the  operational structure}

Summarizing the ideas introduced so far,   an \emph{operational structure} consists of a triple 
\[\Op    =  (  \Transf ,\Outcomes, \Test ) \, ,\]
 where $\Transf$ is a strict symmetric monoidal category, $\Outcomes$ is a collection of sets closed under Cartesian product, and $\Test$ is a strict symmetric monoidal category,  related to $\Transf$ and $\Outcomes$ as described in the previous paragraph.  Intuitively, the operational structure describes \begin{enumerate}
 \item what can be done  (connecting tests) 
 \item what can be observed (outcomes), and
 \item what can happen  (events).  
 \end{enumerate}

\subsection{Probabilistic structure}

%The operational structure just  describes networks of tests and networks of events, without assigning probabilities to the latter.  
The goal of a physical theory is not only to \emph{describe} a class of experiments, but also to \emph{make  predictions} about the outcomes of such experiments.  In the following we show how this can be accomplished by adding a probabilistic structure on top of the operational structure.   

\subsubsection{Assigning probabilities}
An \emph{experiment} consists in sequence of tests that starts from a preparation-test and ends with an observation-test, leaving no open wires, as in the following example  
\begin{align}\label{experiment}
\begin{aligned}
\Qcircuit @C=1em @R=.7em @! R {
 &\prepareC{ \boldsymbol{ \rho}}&\poloFantasmaCn{\rA} \qw &  \gate{\boldsymbol{\tT}}  &\poloFantasmaCn{\rB} \qw   &  \measureD{\bf  m }}
\end{aligned}  
  ~ . \end{align}   
% which describes a preparation-test  $\boldsymbol{\rho}  =\{  \rho_x\}_{x\in\rX}$ for system $\rA$, followed by a test $\boldsymbol{\tT}   =   \{  \tT_y\}_{y\in\rY}$ that transforms system $\rA$ into system $\rB$,  which is in turn followed   by an observation-test  ${\bf m}  =\{  m_y\}_{y\in\rY}$ on system $\rB$.    

If we compose  all the tests involved in an experiment, we  obtain a single test, which transforms  the trivial system into itself.  
%Such a test contains all the events that can take place in the experiment.    
In order to make predictions on the outcomes of the experiment, we need a rule  assigning a probability to the events of such test.  The rule is provided by the probabilistic structure of the theory:  
\begin{definition}[Probabilistic structure]\label{def:probstructure}
Let $\Op$ be an operational structure.  A \emph{probabilistic structure} for  $\Op$ is a map $\Prob:  \Transf (\rI\to \rI)  \rightarrow [0,1]$, which associates a given scalar $s$ to a probability $\Prob(s)$, in accordance to the following two requirements:  
\begin{enumerate}
\item {\em Consistency:}  $\sum_{x\in\rX}    \Prob(s_x) =   1$  for every outcome set $\rX \in  \Outcomes$ and for every test $ \bs s  \in  \Test (\rI\to \rI,  \rX)$
\item {\em Independence:} $\Prob  (s\otimes t)  =  \Prob (s)  \,  \Prob (t)$ for every pair of scalars $s$ and $t$. 
\end{enumerate}
\end{definition}
The consistency requirement guarantees that we can interpret $\Prob(s_x)$ as the probability of the outcome $x \in \rX$. The independence requirement guarantees that experiments that involve only independent tests on two systems give rise to uncorrelated outcomes. As observed by Hardy \cite{hardy11,hardy2013}, independence is equivalent to the requirement that    probabilities can assigned to the outcomes of an experiment in a way that is   independent of the context in which the experiment is performed. Note that the map $\Prob$ needs not be surjective: for example, in a \emph{deterministic} theory the range of  $\Prob$ are only the values 0 and 1.

We are now ready to give the formal definition of OPT:  
\begin{definition}
An \emph{operational-probabilistic theory}   $\Theta$ is a pair  $ (\Op,\Prob)$ consisting of an operational structure  $\Op$ and of a probabilistic structure for $\Op$.  
\end{definition}
\subsubsection{Statistically equivalent events}

Once probabilities are introduced,  it is natural to identify  events that give rise to the same probabilities in all possible circuits.     Precisely, we say that  two events   of type  $\rA\to \rB$, say $\tE$ and $\tE'$, are \emph{statistically equivalent} iff   
\begin{align*}
\Prob\left(\begin{aligned}
\Qcircuit @C=1em @R=.7em @! R {
 &\multiprepareC{1}{ \rho}&\poloFantasmaCn{\rA} \qw &  \gate{{\tE}}  &\poloFantasmaCn{\rB} \qw   &  \multimeasureD{1}{m }\\
  &\pureghost{ \rho}&\poloFantasmaCn{\rR} \qw &  \qw  & \qw   &  \ghost{m }}
\end{aligned}  \quad \right)  \, =\,   \Prob \left( \begin{aligned}
\Qcircuit @C=1em @R=.7em @! R {
 &\multiprepareC{1}{ \rho}&\poloFantasmaCn{\rA} \qw &  \gate{{\tE'}}  &\poloFantasmaCn{\rB} \qw   &  \multimeasureD{1}{m }\\
  &\pureghost{ \rho}&\poloFantasmaCn{\rR} \qw &  \qw  & \qw   &  \ghost{m }}
\end{aligned}  \quad \right)
  \end{align*}   
for every system $\rR$, every preparation-event $\rho  \in \Transf(\rI \to \rA\otimes \rR)$ and every observation-event $m  \in \Transf(\rB\otimes \rR \to \rI)$.   
We denote by $\left[\tE\right]$ the equivalence class of the event $\tE$.

Equivalence classes can be composed in sequence and parallel in the obvious way  
\[[  \tF]  \circ  [  \tE]  : =        [  \tF \circ \tE]  \, ,  \qquad \qquad[\tE]  \otimes [\tF]   :  =   [  \tE\otimes \tF]   \]
  and it is easily verified that both definitions are well-posed. Furthermore,  $[  \tI_\rA]$ and $[\tS_{\rA,\rB}]$ behave like the identity on $\rA$ and the swap between $\rA$ and $\rB$, respectively. 
  As a result, the equivalence classes of events form a strict SMC, which we denote by $\left[ \Transf\right]$.    
  
Similar considerations apply to tests: the equivalence class of a test $  \boldsymbol{  \tE}   =  \{  \tE_x \}_{x\in\rX}$  is defined  as   $ \left[ \boldsymbol{ \tE}\right]   :=  \left\{ \left[ \tE_x\right]  \right \}_{x\in\rX}$ and the sequential/parallel composition of  equivalence classes of tests are induced by the sequential/parallel composition of events: 
\[         [  \boldsymbol{\tF}]  \circ  [  \boldsymbol{\tE}]  :  =  [ \boldsymbol{ \tF} \circ \boldsymbol{\tE}]   \, ,  \qquad \qquad  
[\boldsymbol{\tE}]  \otimes [\boldsymbol{\tF}]  :  =  [ \boldsymbol{ \tE}\otimes  \boldsymbol{\tF}]   \, . \]
Again, the equivalence class of $[  \boldsymbol{\tI}_\rA]$ and $[\boldsymbol{\tS}_{\rA,\rB}]$ behave like the identity and the swap.  As a result, the equivalence classes of tests form a strict SMC, which we denote by $\left[  \Test \right]$.

\subsubsection{The quotient OPT}
The notion of statistical equivalence allowed us to transform the original operational structure $\Op  =  (\Transf,\Outcomes,\Test)$ into a new operational structure $\left[\Op\right]  :=  (\left[  \Transf\right] ,\Outcomes,\left[  \Test\right])  $, which we call the \emph{quotient operational structure}.  The operational structure  $[\Op]$ comes with an obvious probabilistic structure $[\Prob]$, defined as 
\[  \left[  \Prob \right]  (  \, [s]\,  )   :  =   \Prob(s)   \qquad \qquad \forall s\in\Transf(\rI\to \rI)  \, .  \]  
It is indeed immediate to verify  that the consistency and independence conditions in definition \ref{def:probstructure} are satisfied.  
As a result, the original OPT $\Theta   =  (\Op,\Prob)$  has been turned into a new OPT  $[\Theta] :  =  (  [\Op], [\Prob])$, which we call the \emph{quotient OPT}.  Intuitively, the quotient OPT contains all the information that is statistically relevant, disregarding those distinctions that have no consequences for the purpose of making  probabilistic predictions.

In the following we will    focus on quotient OPTs: by default, an OPT will be a \emph{quotient} OPT.  
 Accordingly, we will omit the symbol of equivalence class everywhere and write $\Theta=  (\Op,\Prob)$, assuming that equivalence classes have been already taken from the start.       
This is equivalent to requiring  the following \emph{separation property}  \cite{chiribella14dilation}:       
\begin{definition}\label{ax:separation}
An OPT satisfies the \emph{separation property} iff for every pair of systems $\rA$ and $\rB$ and every pair of events  $\tE$ and $\tE'$ of type $\rA\to \rB$ the condition 
\begin{equation*}
\Prob\left( \begin{aligned}  \Qcircuit @C=.5em @R=0em @!R { 
& \multiprepareC{2}{\rho}    & \qw \poloFantasmaCn{\rA}  &  \gate{\tE}  &  \qw \poloFantasmaCn{\rB}  &\multimeasureD{2}{m} \\
& \pureghost{\rho}   &&&&\pureghost{m}\\
 & \pureghost{\rho}    & \qw \poloFantasmaCn{\rR}  &  \qw &\qw &\ghost{m}} 
\end{aligned}  \quad  \right)  
=
\Prob\left(\begin{aligned}  \Qcircuit @C=.5em @R=0em @!R { & \multiprepareC{2}{\rho}    & \qw \poloFantasmaCn{\rA}  &    \gate{\tE'}  &  \qw \poloFantasmaCn{\rB}  &\multimeasureD{2}{m}  \\
& \pureghost{\rho}   &&&&\pureghost{m}\\
 & \pureghost{\rho}    & \qw \poloFantasmaCn{\rR}  &  \qw&\qw&\ghost{m} } 
\end{aligned} \quad \right)      \quad 
\begin{array}{l}
\forall \rR\in\Sys \\
\forall \rho\in\Transf(\rI  \to \rA\otimes \rR)  \\
 \forall  m\in\Transf(\rB\otimes \rR  \to  \rI)
\end{array}  
 \end{equation*}
 implies $\tE  =\tE'$.  
\end{definition}

In a quotient OPT % we will use a special notation for preparation-events and observation events. 
preparation-events  (respectively, observation-events) will be called \emph{states} (respectively, {\em effects})  and we will use the notation $\St(\rA) :  =  \Transf(\rI\to \rA)$  (respectively, $\Eff(\rA) :  =  \Transf(\rA\to \rI)$).

\subsubsection{Vector space representation of an OPT}

OPTs satisfying the separation property have a convenient  representation in terms of ordered vector spaces and positive maps.   
The construction proceeds in four steps:  

\begin{enumerate}
\item %{\em Step 1.} 
The separation property guarantees that a scalar $s$ can be identified with its probability $\Prob(s)$. Hence, from now on we will omit $\Prob$ and will identify the set of scalars $\Transf(\rI\to \rI)$ with a subset of  the real interval $[0,1]$. 

\item %{\em Step 2.} 
By the separation property, a state $\rho  \in  \St(\rA)$ can be identified with the real-valued  function    $\hat \rho: \Eff(\rA)  \to \R $ defined by 
\[   \widehat \rho  (  m)   :  =    \begin{aligned}
    \Qcircuit @C=1em @R=.7em @! R {
      &\prepareC{\rho}&\poloFantasmaCn{\rA}\qw  &\measureD{m} }  
      \end{aligned} \]
     (indeed, one has $\rho  =  \sigma$ if and only if $  \widehat \rho  =  \widehat{ \sigma}$).  
  Since real-valued functions form a vector space, we can define the vector (sub)space spanned by the states of system $\rA$ as  
\[  \Str  (\rA)   :   =  \Span_\R  \left\{   \rho  ~|~ \rho  \in  \St (\rA)   \right\}  \, .\]
Limiting ourselves to linear combination with positive coefficients we obtain the proper cone   $\St_+ (\rA)$, which turns $\Str(\rA)$ into an ordered vector space.

\item %{\em Step 3.}  
Every effect  $m\in\Eff  (\rA)$  defines a \emph{linear} function $\widehat m:  \St_\R (\rA)  \to \R$, via the relation 
\[  \widehat m  \left(   \sum_i  c_i\,   \rho_i  \right)  :  = \sum_i  c_i\,       \begin{aligned}
    \Qcircuit @C=1em @R=.7em @! R {
      &\prepareC{\rho}&\poloFantasmaCn{\rA}\qw  &\measureD{m_i} }    
      \end{aligned}  \, ,   \qquad  \quad\forall \{c_i\}  \subset \R \,  , \quad \forall \{\rho_i\}  \subset \St(\rA)  \, .    \]
It is immediate to see that the definition is well-posed, namely $\widehat m  \left(   \sum_i  c_i\,   \rho_i  \right)  =  \widehat m  \left(   \sum_j  c'_j\,   \rho'_j  \right) $
 whenever $ \sum_i  c_i\,   \rho_i    =   \sum_j  c'_j\,   \rho'_j$.    Again, the effect $m$ can be identified with the linear function $\widehat m$ thanks to the separation property. 
 Taking linear combinations of effects we obtain the vector space 
\[  \Effr  (\rA)   :   =  \Span_\R  \left\{   m  ~|~ m  \in  \Eff (\rA)   \right\}   \, ,\] 
while restricting  to positive linear combinations we obtain  the  proper cone $\Eff_+(\rA)$.   As a result, also $\Eff_\R(\rA)$ is  an ordered vector space.

\item 
%{\em Step 4.}   
Every event  $\tE$  of type $\rA\to \rB$ induces a linear map $\widehat  {\tE}  :  \St_\R (\rA) \to \St_\R  (\rB)$, via the definition  
\[  \widehat {\tE}  \left(   \sum_i  c_i\,   \rho_i  \right)  :  = \sum_i  c_i\,     \left(   \tE\circ \rho_i   \right)  \, ,   \qquad \qquad \qquad\forall \{c_i\}  \subset \R \,  , \quad \forall \{\rho_i\}  \subset \St(\rA)  \, .    \]  
 Again, it is not hard to see that the definition is well-posed, namely that $\widehat {\tE}  \left(   \sum_i  c_i\,   \rho_i  \right)  =  \widehat {\tE}  \left(   \sum_j  c'_j\,   \rho'_j  \right) $
 whenever $ \sum_i  c_i\,   \rho_i    =   \sum_j  c'_j\,   \rho'_j$.        Note that the map $\widehat{\tE}$ is not only linear, but also \emph{positive}:  indeed, it sends elements of the cone $\St_+(\rA)$ to elements of the cone $\St_+(\rB)$.   We call $\widehat{\tE}$ the \emph{state change} associated to $\tE$.
 
 %At this point, there is an important caveat: 
 
 % 
 \end{enumerate}

 %This is indeed the case for quantum theory on real Hilbert spaces.  

At this point, a few remarks are in order:

\begin{enumerate}
\item \emph{Linearity vs convexity.}    Traditionally, the linearity of state changes has been argued from the assumption that the state space $\St(\rA)$ is convex. However, our argument shows that such assumption is \emph{not} needed: the probabilistic structure alone suffices to define the linear map $\widehat{\tE}$.
\item {\em Finite vs infinite dimensional systems.}  For a given system $\rA$, we define $D_\rA$ to be the dimension of the  vector space $\St_\R(\rA)$ and    we say that  system  $\rA$ is \emph{finite dimensional}  if $D_\rA$ is finite.  
For finite systems, one has the equality   $\Eff_\R  (\rA)  =  \St_\R(\rA)^*$, where $  \St_\R(\rA)^*$ is the vector space of all linear functionals on  $\St_\R(\rA)$.  For infinite dimensional systems, such an equality may not hold.    
%As a result, finite dimensional systems can be characterized as those systems 
%Operationally,  one has that $D_\rA-1$ is the minimum number of binary observation-tests needed to characterize the state of the system completely (in the statistical sense:  once the outcome probabilities of these tests are known, the state is completely identified).     

\item {\em The no-restriction hypothesis.}   Since effects are  identified with positive linear functions, one has the inclusion $\Eff_+ (\rA)  \subseteq  \St_+(\rA)^*$, where $\St_+(\rA)^*$ denotes the dual cone of $\St_+(\rA)$  
\begin{align}
\St_+(\rA)^*  :  =  \{  \,   m    \in  \St_\R(\rA)^*  ~|~    m(  \rho)   \ge  0   \quad \forall \rho\in\St_+  (\rA)\} \, .
\end{align}
Even for finite dimensional systems, the inclusion $\Eff_+ (\rA)  \subseteq  \St_+(\rA)^*$ may not be an equality.   The  assumption $\Eff_+ (\rA)  = \St_+(\rA)^*$  is known as \emph{No-Restriction Hypothesis}  \cite{puri}.  We stress that such an assumption is \emph{not} made in our derivation.  
\item {\em Transformations vs linear maps.}  Unlike in the case of states and effects, the correspondence between the transformation $\tE$ and the linear map $\widehat{\tE}$ may not be one-to-one. The reason for this is that the difference between two transformations $\tE$ and $\tE'$  may  show up  when one applies them locally on a part of a composite system: one can have      $\widehat  {\tE\otimes \tI_\rR}   \not   =   \widehat  {\tE'\otimes \tI_\rR} $ for some $\rR\in\Sys$ even if $\widehat{\tE}  =  \widehat{\tE'}$.    This problem disappears if one assumes the axiom of Local Tomography, as we will see  later in this chapter.     In the lack of Local Tomography,  however, the transformation $\tE$ can still be identified with a linear map:  for this purpose, one can choose the  linear map    $\widehat{\tE}_{\oplus}$ defined by \cite{chiribella14dilation}
\begin{align}\label{eoplus}  \widehat {\tE}_\oplus  :  =  \bigoplus_{\rR\in\Sys}       \,  \widehat {\tE\otimes \tI_\rR} \, .   \end{align}
The map  $\widehat{\tE}_\oplus$   transforms elements of the (infinite-dimensional) vector space $ \St_{\R,\oplus}  (\rA)  :  =  \bigoplus_{\rR\in\Sys}   \St_\R  (\rA\otimes \rR) $ into elements of the  (infinite-dimensional) vector space $ \St_{\R,\oplus}  (\rB)  :  =  \bigoplus_{\rR\in\Sys}   \St_\R  (\rB\otimes \rR) $.       Then, the separation property guarantees that the correspondence between $\tE$ and $\widehat{\tE}_\oplus$ is one-to-one. 
 \item {\em The vector space of transformations.}  So far we have defined the vector spaces of states and effects.  A vector space of  transformations can be defined using the one-to-one correspondence with the linear maps in   Eq. (\ref{eoplus}) and setting 
 \begin{align}\label{transfR}
   \Transf_\R  (\rA\to \rB)  :  =    \Span_\R  \{   \Transf(\rA\to \rB)\}  \, .
   \end{align} 
   Again, a proper cone   $\Transf_+  (\rA\to \rB)$ can be defined by restricting the attention to linear combinations with positive coefficients. Note that, in general, the vector space   $\Transf_\R  (\rA\to \rB)  $ and the cone  $\Transf_+  (\rA\to \rB)$ can be infinite-dimensional \emph{even if both systems $\rA$ and $\rB$ are  finite dimensional}.  However, this is not the case when the theory satisfies the Local Tomography. 
   \end{enumerate}
 
 \subsubsection{Closure under coarse-graining}

A key notion that comes with the probabilistic structure is the notion of \emph{coarse-graining}:  given a test $\boldsymbol{\tT}  =\{\tT_x\}_{x\in \rX}$, one can decide to identify some outcomes, thus obtaining another, coarse-grained test.  Mathematically, a coarse-graining is defined by  partitioning  the outcome set $\rX$ into mutually disjoint subsets $\{  \rX_y\}_{y\in\rY}$.      Relative to such partition, the coarse-graining of the  test  $\boldsymbol {\tT}$ is the test $\boldsymbol{\tT}'  = \{\tT'_y\}_{y\in\rY}$ defined by\footnote{Note that  the summation is well-defined  thanks to the  vector space structure of $\Transf_\R (\rA\to \rB)$.  }
\begin{align}\label{coarse}
\tT'_y  :  =  \sum_{x\in\rX_y}   \tT_x  \, , 
\end{align}
setting $\tT_y'   =   0$ for $\rX_y  =  \emptyset$, where $0$ is the zero element of the vector space $\Transf_\R (\rA\to \rB)$.   Note that, by calling $\bs \tT'$ a \emph{test} we have  implicitly made two assumptions, namely that 
\begin{enumerate}
\item the set $\rY$ belongs to $\Outcomes$
\item the collection  $\{\tT'_y\}_{y\in\rY}\subset \Transf_\R(\rA\to \rB)$ belongs to $\Test(\rA\to \rB,\rY)$.
\end{enumerate}
% Demanding that these two conditions are satisfied  means enriching the original OPT  in natural way arising  from  the probabilistic structure.  
 From now on, we will  require that our OPT is \emph{closed under coarse-graning}  meaning that the above conditions are satisfied.
 %  for every pair of systems $\rA$ and $\rB$, every outcome set $\rX $, every disjoint partition  $\{  \rX_y\}_{y\in\rY}$ of $\rX$, and every test $\boldsymbol{\tT}\in\Test(\rA\to \rB,\rX)$. 
 
By coarse-graining over all outcomes of a test $\bs \tT  \in  \Test(\rA\to \rB,  \rX)$ one obtains a deterministic test, identified with the deterministic transformation $\tT  :  =  \sum_{x\in\rX}   \,  \tT_x \in \Det\Transf(\rA\to \rB)$.   In particular, when a preparation test $\bs \rho  \in\Test(\rI\to \rA,\rX)$ satisfies  $\sum_{x\in\rX}  \,  \rho_x  =  \rho$ we say that the test $\bs \rho$ is an \emph{ensemble decomposition of $\rho$}.

\subsubsection{Summary of the OPT framework}   

Let us sum up the main points  discussed so far.  
We defined an OPT   as a pair  $\Theta  =  (  \Op,\Prob)$,  consisting of an operational structure $\Op  =  (  \Transf,\Outcomes, \Test)$ and of a probabilistic structure $\Prob$ that assigns probabilities to  scalars. 
 We restricted our attention to  OPTs that satisfy the Separation Property (definition \ref{ax:separation}), which implies that one can identify  scalars with probabilities, states with elements of suitable vector spaces, and  effects with linear functionals over them.  Transformations with nontrivial input and output induce linear maps on the corresponding state spaces.    Finally, in agreement with the probabilistic interpretation, we demanded that the theory $\Theta$ is closed under coarse-graining.

\section{Background of  the quantum reconstruction}\label{sec:background}
In this section we provide some background that will be useful for our reconstruction of quantum theory. 
We start by reviewing three standing assumptions: finite-dimensionality,  non-determinism, and closure under operational limits. 
 We will then review the operational tasks that motivate our axioms. 
 \subsection{Standing assumptions}
Here we introduce three standing assumptions that will be made in the rest of the   chapter.  These assumptions are common to all recent axiomatizations of quantum theory, and could also be even incorporated in the OPT framework. We keep them separate from the rest, both for clarity of presentation and for the sake of maintaining the OPT framework as flexible as possible.   The assumptions are the following:
\begin{enumerate}

\item {\em Finite dimensionality.}  We restrict our attention to finite systems,~i.~e.~systems with finite dimensional state spaces. Operationally, this means that  the state of every system can be identified from the statistics of a \emph{finite number of finite-outcome measurements}.   Of course, the implicit assumption here is that finite systems exist  and form a sub-theory of  our theory, meaning that the operational structure $\Op$ contains a non-trivial substructure $\Finite\Op$, consisting of transformations, outcome sets, and tests involving only finite systems.

\item {\em Non-determinism.}  While the OPT framework accommodates a variety of theories,  here we focus on  OPTs that are non-deterministic, meaning that  there exists at least one experiment for which the outcome is not determined a priori.  Mathematically, this means that the range of the probability function $\Prob$ is not just $\{0,1\}$.  Note that  non-determinism  is a weaker assumption than convexity of the state spaces: there exist indeed examples of theories---such as Spekkens' toy theory \cite{SpekkensToy}---that are non-deterministic and yet do not have convex state spaces.

\item {\em Closure under operational limits.}   Suppose that $( \tT_n )_{n\in\N}$ is a sequence of transformations  of type $\rA  \to \rB$ and that $\tT$ is an element of the vector space $ \Transf_\R  (\rA\to \rB)$ such that   
\begin{equation*}
\lim_{n\to\infty}  \begin{aligned}  \Qcircuit @C=.5em @R=0em @!R { 
& \multiprepareC{2}{\rho}    & \qw \poloFantasmaCn{\rA}  &  \gate{\tT_n}  &  \qw \poloFantasmaCn{\rB}  &\multimeasureD{2}{m} \\
& \pureghost{\rho}   &&&&\pureghost{m}\\
 & \pureghost{\rho}    & \qw \poloFantasmaCn{\rR}  &  \qw &\qw &\ghost{m}} 
\end{aligned}  
=
\begin{aligned}  \Qcircuit @C=.5em @R=0em @!R { & \multiprepareC{2}{\rho}    & \qw \poloFantasmaCn{\rA}  &    \gate{\tT}  &  \qw \poloFantasmaCn{\rB}  &\multimeasureD{2}{m}  \\
& \pureghost{\rho}   &&&&\pureghost{m}\\
 & \pureghost{\rho}    & \qw \poloFantasmaCn{\rR}  &  \qw&\qw&\ghost{m} } 
\end{aligned}       \qquad 
\begin{array}{l}
\forall \rR\in\Sys \\
\forall \rho\in\Transf(\rI  \to \rA\otimes \rR)  \\
 \forall  m\in\Transf(\rB\otimes \rR  \to  \rI)  \, , 
\end{array}  
 \end{equation*}
 meaning that the probability of every experiment involving $\tT_n$ converges to the probability of an hypothetical  experiment  involving $\tT$. 
  When this is the case, we assume that $\tT$ belongs to $\Transf(\rA\to \rB)$.   Operationally,  one can think of the sequence $(\tT_n)_{n\in\N}$ as a \emph{limit procedure} to implement the transformation $\tT$.

\end{enumerate}   

\subsection{Basic operational tasks} 

We now give a brief list of the operational notions on which our axioms are based.  
\subsubsection{Signalling}

When a number of devices  are connected into a network, it is natural to ask whether one node of the network can  signal to another.   
  For example, given the experiment
  \begin{equation}\label{genericcircuit}
  \begin{aligned}  \Qcircuit @C=.5em @R=0em @!R { 
& \multiprepareC{2}{\boldsymbol{\rho}}    & \qw \poloFantasmaCn{\rA}  &  \gate{\boldsymbol{\tT}}  &  \qw \poloFantasmaCn{\rB}  & \multigate{2}{\boldsymbol{\tE}}  & \qw \poloFantasmaCn{\rC} &   \gate{\boldsymbol{\tS}}   &  \qw \poloFantasmaCn{\rD} &   \multimeasureD{2}{\boldsymbol{m}}   
\\  % &\pureghost{supercalifragil}   & { \boldsymbol{\rho}  =  \{\rho_{x_1}\}_{x_1\in\rX_1}} \, ,& &\\
& \pureghost{\boldsymbol{\rho}}   &&&    &  &  &&& &\pureghost{m}  \\
%& \boldsymbol{\tT}  =  \{\tT_{x_2}\}_{x_2\in\rX_2}  \, ,& \pureghost{supercalifragi}&   \boldsymbol{\tE}  =  \{\tE_{x_3}\}_{x_3\in\rX_3}  \, ,\\
 & \pureghost{\boldsymbol{\rho}}    & \qw \poloFantasmaCn{\rR}  &  \qw &\qw &\ghost{\boldsymbol{\tE} }  & \qw \poloFantasmaCn{\rS} &  \qw &\qw  & \ghost{\boldsymbol{m}}  
 %&& \boldsymbol{\tS}  =  \{\tS_{x_4}\}_{x_4\in\rX_4} \, ,&\pureghost{supercalifragi}  & ~\boldsymbol{m}  =  \{m_{x_5}\}_{x_5\in\rX_5} 
 \, , } 
\end{aligned}  
\end{equation}
one can ask whether the choice of the test $\boldsymbol{\tT}$ can influence the outcome of the test $\boldsymbol{\tS}$.   Precisely, the question is whether  or not the marginal probability distribution for the outcomes of  $\bs \tS$ (obtained by summing over the outcomes of all the other tests in the network) depends on $\bs \tT$.   
 %In order to answer the question, one has to  keep  fixed all the tests in the circuit except for  $\boldsymbol{\tT}$, and to check whether or not a variation of $\boldsymbol{\tT}$ leads to a variation of the outcome probabilities for   $\boldsymbol{\tS}$.  
% Explicitly, the probability distribution for such outcome is given by
 %\[     p(   x_4   |   \boldsymbol{\tT})   :  =     
 %\sum_{x_1  \in \rX_1}   \sum_{x_2  \in \rX_2}    \sum_{x_3  \in \rX_3}  \sum_{x_5  \in \rX_5} \,  (  m_{x_5}|
 % ( \tS_{x_4}  \otimes \tI_\rS)   \,  \tE_{x_3}  \,  (\tT_{x_2}  \otimes  \tI_\rR)   \,  |\rho_{x_1}  ) \, .     \]
Denoting the marginal probability distribution by $p(x|  \bs \tT)$,  $x\in\rX$,  we say that the node occupied by the test $\boldsymbol{\tT} $ \emph{does not signal} to the node occupied by the test $\boldsymbol{\tS}$  iff 
 \[  p(x| \boldsymbol{\tT}_0  )   =  p(x  |\boldsymbol{\tT}_1)  \qquad \qquad \forall   x  \in \rX  %\, , \,  \forall \rY_0, \rY_1 \in \Outcomes, \, , \,  \forall \boldsymbol{\tT}  \in  \Test(\rA\to \rB,\rY_0),\, \forall \boldsymbol{\tT'}\in  \Test(\rA\to \rB,\rY_1) 
 \, ,\] 
for every possible choice of tests $\tT_0$ and $\tT_1$. 
Similarly, one can ask whether the node occupied by the test $\boldsymbol{\tS}$ can signal to the node occupied by the test $\boldsymbol{\tT}$.  
%,~i.~e.~whether or not the probability distribution    $p(y|  \bs \tS)$,  $y\in\rY$   depends on $\boldsymbol{\tS}$. 
  Now,  note that the test  $\boldsymbol{\tS}$ is performed \emph{after} the test $ \boldsymbol{\tT}$: 
  if  the node occupied by $\boldsymbol{\tS}$ can signal to the node occupied by  $\boldsymbol{\tT}  $  we say that the circuit of Eq. (\ref{genericcircuit}) allows for \emph{signalling from the future to the past}. 
 % When this is the case for some circuit, we say that the OPT allows for signalling from the future to the past.  

%These circuits are a special case of the circuits in Eq. (\ref{genericcircuit}), obtained by setting $\rB\equiv\rC$, $\rA\equiv\rR\equiv \rD\equiv \rS  \equiv \rI$,   $\boldsymbol{\rho}  \equiv \boldsymbol{\tI}_\rI $,   $\boldsymbol{\tE} \equiv  \boldsymbol{\tI}_\rB   $, and $\boldsymbol{m}  \equiv \boldsymbol{\tI}_\rI$.  The reduction to these circuits is immediate, because 
%The impossibility of signalling from the future to the past will be the first axiom in our list. 

\subsubsection{Collecting side information}
 
 Suppose that the test $\bs \tT =  \{   \tT_x\}_{x\in\rX}$ is obtained from the test $\bs\tT'  =  \{   \tT_z\}_{z\in\rZ}$ via coarse-graining, namely  
 \[\tT_x  =   \sum_{z\in\rZ_x}   \,  \tT_z'   \qquad \forall x\in\rX \, , \] 
 where  $\{\rZ_x\}_{x\in\rX}$ is a partition of $\rZ$ into disjoint subsets.  In this case we say that $\bs \tT'$ \emph{refines}  $\bs \tT$.  Now, it is convenient to relabel the outcomes of    $\bs \tT'$ as $z  =  (x,y)$, with $x\in \rX$ and $y\in \rZ_x $, and to write $\tT'_{x,y}$ in place of $\tT'_z$.    In this way, we can think of the random variable $y$ as a  \emph{side information}, which is not accessible to the agent  Alice  performing the test $\bs \tT$, but may be accessible to some other agent  Eve.  
 This picture is particularly relevant to cryptographic scenarios, wherein Eve could be  an eavesdropper  attempting  to collect as much information as possible. 
 In all such scenarios, a special role is played by those transformations that do not leak  any useful side information.   We call such transformations \emph{pure}:  
\begin{definition}    
  We say that a transformation $\tE$ is \emph{pure}\footnote{In previous works, we used different names for transformations that do not allow for side information:  in Refs. \cite{puri,deri}  they were called \emph{atomic}, while in the popularized version of  Ref.  \cite{chiribella2012quantum} they were called \emph{fine-grained}.  We apologize with our readers for the changes of terminology,  due to an ongoing search for the word that best captures this  operational concept.  In this chapter, we adopted  the word \emph{pure}, because \emph{i)} this term is the standard one in the case of states and \emph{ii)} using the same term for transformations should hopefully ease the reading.   Still, a warning is in order:    when the set of transformations $\Transf(\rA\to \rB)$  is convex, the pure transformations $\Pur\Transf(\rA\to \rB)$ \emph{may not coincide} with the extreme points of $\Transf(\rA\to \rB)$.     For example, in quantum theory the identity effect $ I_\rA$  is an extreme point of the set of effects, but is not pure in the sense of our definition because it can be decomposed~e.~g.~as $I_\rA  =  \sum_{n=1}^{d_\rA}  \,  P_n$, where the effects $ \{P_n =  |n\>\<  n|~|~  n=1,\dots, \,  d_\rA \}$ represent a projective measurement on some orthonormal basis $\{   |n\>~|~  n=  1,\dots \, , d_\rA  \}$. }   
  iff for every test $\bs  \tT$ containing $\tE$ and for every  test $\bs\tT'$ refining $\bs \tT$ one has 
 \begin{align}\label{purity}  \tT'_{x_0  ,y}   \, = \,    p_y   ~ \tT_{x_0}  \, ,  
 \end{align}
 where $x_0$ is the outcome such that $\tT_{x_0}   =  \tE$ and $\{p_y\}$ is a probability distribution. 
  \end{definition}

Informally, the purity condition  (\ref{purity}) states that the side information  possessed  by Eve   is uncorrelated with the transformation  $\tE$ taking place in Alice's laboratory.    We denote the set of pure transformations of type $\rA\to \rB$ by $\Pur\Transf(\rA\to \rB)$.  In the special case of transformations with trivial input we will  use the notation $\Purs\St(\rA)$  (respectively, $\Pur\Eff (\rA)$), referring to \emph{pure states} (respectively, \emph{pure effects}).  An \emph{pure test} is a test consisting of pure transformations.

 Transformations that are not necessarily pure will be called \emph{mixed}.  
  Among the mixed transformations, the ones that are in the interior of the cone $\Transf_+(\rA\to \rB)$ play an important role.  
 They are defined as follows:  
  \begin{definition}
  A transformation $\tE \in\Transf(\rA\to \rB)$ is called \emph{internal} iff for every transformation $\tF \in\Transf(\rA\to \rB)$ there exists a transformation $\tG$ and  a scaling constant $\lambda >  0$    such that  
    \begin{enumerate}
  \item $\tE =  \lambda \,  \tF  +  \tG $
  \item $\lambda  \,  \tF$ and $\tG$ coexist in a test \footnote{Note that, in principle, our definition of ``internal transformations"  may not include all the transformations in the interior of the cone, because the   $\lambda \,  \tF$ and $\tG$ may fail to coexist in a test.   However, this annoying discrepancy disappears under the mild assumption that the set of transformations is convex.  Later, we will justify this assumption on the basis of the Causality axiom.     }.  
  \end{enumerate} 
  \end{definition}
Roughly speaking, an internal transformation is compatible with the occurrence of any other transformation of the same type. 
Internal transformations with trivial input (output) will be called \emph{internal states}  (\emph{internal effects}).

  \subsubsection{State tomography}

The task of state tomography consists in identifying the state of a system from the statistics of a restricted set of observations.  
  Suppose that an experimenter is able to perform a set of observation-tests   and let   $\rM$ be the set of all effects appearing in such tests.   
  \begin{definition}
  We say that the effects in  $\rM$ are \emph{tomographically complete} for system $\rA$ iff, for every pair of states $\rho$ and $\rho'$ of system $\rA$, one has  the implication
  \begin{align*}       
  &  \begin{aligned}
    \Qcircuit @C=1em @R=.7em @! R {
      &\prepareC{\rho^{\phantom{\prime}}_{\phantom x}}&\poloFantasmaCn{\rA}\qw  &\measureD{m^{\phantom{\prime}}_{\phantom x}} }  
      %\end{aligned}
         \quad =  
         %  \begin{aligned}
    \Qcircuit @C=1em @R=.7em @! R {
      &\prepareC{\rho'_{\phantom x}}&\poloFantasmaCn{\rA}\qw  &\measureD{m^{\phantom{\prime}}_{\phantom x}} }
      \end{aligned}  
       \qquad  \forall  m\in  \rM  \\    \\ 
       & \qquad  \qquad \Longrightarrow \qquad            
      \begin{aligned}
    \Qcircuit @C=1em @R=.7em @! R {
      &\prepareC{\rho'}&\poloFantasmaCn{\rA}\qw  &\qw    }  
      \end{aligned}   \quad   =   \begin{aligned}
    \Qcircuit @C=1em @R=.7em @! R {
      &\prepareC{\rho^{\phantom{\prime}} }&\poloFantasmaCn{\rA}\qw  &\qw    }  
      \end{aligned}    
       ~ .
      \end{align*} 
      \end{definition}
In the contrapositive: if two states are different, then the difference can be detected from the statistics of some effect   in  $\rM$.

Let us consider state tomography for composite systems.
Suppose that two experimenters  Alice and Bob  perform measurements on two systems $\rA$ and $\rB$, respectively, and that Alice (Bob) is able to perform the  set of measurements with effects $\rM$ ($\rN$).  Then, by coordinating their choices of measurements and by communicating the outcomes to each other, Alice and Bob can observe the statistics of all  product measurements. Hence, their set of measurement effects will be 
\[   \rM\otimes  \rN :  =   \{    {\bf m} \otimes {\bf n}  ~|~   {\bf m} \in  \rM  \, , \quad {\bf n}  \in  \rN\} \, .\]
Now the question is:  is there a choice of measurement effects $\rM$ and $\rN$ such that the set  $\rM \otimes  \rN$ tomographically complete?  
 In the affirmative case, we say that system $\rA\otimes \rB$ \emph{allows for local tomography}: 
 \begin{definition}  System $\rA\otimes \rB$ \emph{allows for local tomography}  iff, for every pair of states $\rho  , \rho'  \in\St(\rA\otimes \rB)$, one has the implication 
 \begin{align}
&   \begin{aligned}
    \Qcircuit @C=1em @R=.7em @! R {\multiprepareC{1}{\rho}&\poloFantasmaCn{\rA}\qw&\measureD{a}\\
      \pureghost{\rho}&\poloFantasmaCn{\rB}\qw&\measureD{b}}
  \end{aligned}  = 
  \begin{aligned}    
    \Qcircuit @C=1em @R=.7em @! R {\multiprepareC{1}{\rho'}&\poloFantasmaCn{\rA}\qw&\measureD{a}\\
      \pureghost{\rho'}&\poloFantasmaCn{\rB}\qw&\measureD{b}}
  \end{aligned}  \qquad  
  \begin{aligned}
  \begin{array}{l}
  \forall a \in\Eff (\rA)  \, ,\\
   \forall b \in  \Eff (\rB)^{\phantom{A^{A^{A^A}}}}
  \end{array}
  \end{aligned}\\   \nonumber \\
   &\quad    \Longrightarrow \qquad
  \begin{aligned}\Qcircuit @C=1em @R=.7em @! R {
      \multiprepareC{1}{\rho}&\poloFantasmaCn{\rA}\qw&\qw\\
      \pureghost{\rho}&\poloFantasmaCn{\rB}\qw&\qw}
  \end{aligned} \quad   =  \quad   \begin{aligned}
    \Qcircuit @C=1em @R=.7em @!  R {\multiprepareC{1}{\rho'}&\poloFantasmaCn{\rA}\qw&\qw\\
      \pureghost{\rho'}&\poloFantasmaCn{\rB}\qw&\qw}
  \end{aligned}
\end{align}
\end{definition}
 More generally, we have the following 
 \begin{definition}
An $K$-partite  system $\rA  =  \bigotimes_{k=1}^K  \rA_k$ \emph{allows for local tomography} iff for every $k \in\{1,\dots, K\}$ there exists a set of measurement effects $\rM_k$  on system $\rA_k$ such that the set $ \bigotimes_{k=1}^K  \rM_k $ is tomographically complete. 
 \end{definition}
For a given OPT, it is easy to see that the following conditions are equivalent:  
\begin{enumerate}
\item every multipartite system allows for local tomography
\item every bipartite system allows for local tomography.
 \end{enumerate}
In other words, the possibility of local tomography for arbitrary composite systems  can be established by just checking  bipartite systems.

\subsubsection{State discrimination}

The task of state discrimination can be presented as a game featuring a player and a referee. The referee prepares a physical system $\rA$ in a state $\rho_x$, belonging to some set $\{  \rho_x ~|~  x\in\rX\}$ known to the player.   
%For simplicity we assume that the states in the set are deterministic---i.e. that $\bs \rho_x : = \{\rho_x\}$ is a preparation-test for every $x\in\rX$.   
The player is asked to guess the label $x$.  In order to do that, she performs a measurement ${\bf m}$ with outcomes  in $\rX$:   
  upon finding the outcome $x'$, she  will guess that the state was $\rho_{x'}$.   
If the player guesses right all the times, we say that the states are perfectly distinguishable: 
\begin{definition}
The states $\{  \rho_x ~|~  x\in\rX\}$ are {\em perfectly distinguishable}  iff there exists a measurement ${\bf m} $  such that 
\[(m_{x}|  \rho_{x'})  =  \delta_{x,x'} \qquad \qquad \forall x,x'\in\rX \, .  \]   
When this is the case, we say that $\bs m$ is a \emph{discriminating measurement}. 
\end{definition}
Note that, in order to be perfectly distinguishable, the states must be 
\begin{enumerate}
\item \emph{normalized}, namely $\|  \rho_x \|   =  1  \, \forall x\in \rX$, where $\|  \cdot  \|$ is the operational norm  \cite{puri} given by   $ \|  \rho  \|    =  \sup_{a\in\Eff(\rA)} \,  (a|\rho)$ 
\item \emph{non-internal}:  indeed, if a state  $\rho_{x'}$ is internal, then   $(m_x|\rho_{x'})  = 0$ implies $m_x  =  0$, in contradiction with the condition $(m_x|\rho_x)=1$.
\end{enumerate}
%Perfectly distinguishable states can be used to communicate  messages reliably: a sender, Alice, could encode the message   $x\in\rX$ into the state $\rho_x$ and transmit the system to a receiver, Bob, who could then decode without error using the measurement $\bf m$.   
Note that \emph{a priori} an OPT may not have any distinguishable states at all.  However,  the existence of distinguishable states is essential if we want our theory to include classical computation and classical information theory.

\subsubsection{Ideal compression}

A  preparation-test   $\bs \rho  \in \Test (\rI\to \rA,\rX) $   can be thought as describing a \emph{source of information}. 
%producing a number of alternative messages.  
     An interesting question is how well such information can be transferred from the original system to another physical support, say system $\rB$.  An  {\em encoding}    of the preparation-test $\bs \rho$  is a deterministic transformation   $\tE  \in \Det \Transf(\rA\to \rB)$, which transforms  $\bs \rho$ into a new preparation-test $\bs  \rho'  :  =    \{\tE     \circ \rho_x\}_{x\in\rX}$.   The states $\{\tE\circ \rho_x~|~ x\in\rX\}$ are called  \emph{codewords}.  

The ideal property of an encoding is to be lossless, in the following sense:
   \begin{definition}
An encoding $\tE  \in \Det \Transf(\rA\to \rB)$ is  \emph{lossless for the preparation-test} $\bs \rho\in\Test(\rI\to \rA,\rX) $ iff there exists a  deterministic transformation $\tD  \in\Det\Transf(\rB\to \rA)$, called the \emph{decoding}, such that 
\begin{align}\label{lossless}
\begin{aligned}
\Qcircuit @C=1em @R=.7em @! R {
 &\prepareC{ { \rho_x}}&\poloFantasmaCn{\rA} \qw &  \gate{{\tE}}  &\poloFantasmaCn{\rB} \qw   &  \gate{\tD} & \poloFantasmaCn{\rA} \qw  & \qw     }
\end{aligned}    =   \begin{aligned}
\Qcircuit @C=1em @R=.7em @! R {
 &\prepareC{ { \rho_x}}&\poloFantasmaCn{\rA} \qw &\qw  }
\end{aligned}     \qquad \forall x\in\rX \, .
\end{align}   
We say that
\begin{itemize}
\item 
$\tE$ is a \emph{lossless encoding for the state $\rho  \in  \Det\St(\rA)$} iff $\tE$ is a lossless encoding for every ensemble decomposition of $\rho$.
% all preparation-tests  $\bs \rho  \in\Test(\rI\to \rA,\rX)$, with arbitrary outcome set $\rX$, satisfying $\sum_{x\in\rX}  \rho_x  =  \rho$
\item $\tE$ a \emph{lossless encoding of system $\rA$ into system $\rB$} iff $\tE$ is a lossless encoding for all states $\rho\in\Det\St(\rA)$. 
\end{itemize}
\end{definition}
The notion of encoding offers an operational  way to compare the size of different systems: naturally, we can say that system $\rA$ is \emph{no larger}
 than system $\rB$ iff there exists a lossless encoding of $\rA$ into $\rB$.    
  
 Among all possible encodings, we now consider the compressions: 
 \begin{definition}
A \emph{compression} of system $\rA$ into system $\rB$ is an encoding $\tE  \in\Det\Transf(\rA\to \rB)$  where $\rB$ is no larger than $\rA$.
\end{definition}

How much can we compress a given state?    The ultimate limit to compression  is when \emph{every} state of system $\rB$ is proportional to a codeword,~i.~e.~when every state $\sigma\in\St(\rB)$ can be written as  $\sigma   =    \lambda \,  \tE   \rho_{x_0}$,  
for some scaling constant $\lambda \ge 0$ and some state $\rho_{x_0}$ belonging to some ensemble decomposition of $\rho$. 
% preparation-test $\bs \rho =  \{\rho_x\}_{x\in\rX}$ satisfying $\sum_{x\in\rX}  \rho_x  =  \rho$.   
   When this is the case, we say that the compression  $\tE$ is \emph{maximally efficient}. Summing up, we have the following  
   \begin{definition}
A transformation $\tE\in\Det\Transf(\rA\to \rB)$ is \emph{ideal compression of the state $\rho\in\Det\St(\rA)$}  iff  it is lossless and maximally efficient.    \end{definition}

\subsubsection{Simulating preparations}

A state can be prepared in many different ways. For example, a state $\rho_\rA$  could be prepared by a circuit that involves many auxiliary systems, which interact with $\rA$ and are finally discarded.  We refer to these systems as the \emph{environment} and describe them collectively as a single system $\rE$.   Assuming that the system and the environment are initially uncorrelated, the fact that the circuit prepares  the state $\rho_\rA$ is expressed by the diagram
 \begin{equation}\label{simulation}
   \begin{aligned}  \Qcircuit @C=.5em @R=0em @!R { 
& \prepareC{{\rho_{0}} }    &\qw & \qw \poloFantasmaCn{\rA} &\qw  & \multigate{2}{{\tU}}  & \qw &\qw \poloFantasmaCn{\rA} &   \qw   &   {\qquad} &  =  &  {\phantom{supercalifragi}}  & \prepareC{{\rho_{A}} }    & \qw \poloFantasmaCn{\rA} &\qw  &\qw 
\\
& \pureghost{{\rho}} &  & &&  \pureghost{{\tU} }  &&&  &&&&&&&\\
%&& & \pureghost{supercalifragi}&   \\
 & \prepareC{{\eta_0}}    &\qw & \qw \poloFantasmaCn{\rE}  &\qw& \ghost{{\tU} }  & \qw & \qw \poloFantasmaCn{\rE} & \measureD{e} &&&&&&&
 %&& \boldsymbol{\tT}  =  \{\tT_{x_3}\}_{x_4\in\rX_4} \, ,&\pureghost{supercalifragi}  & ~\boldsymbol{m}  =  \{m_{x_5}\}_{x_5\in\rX_5} 
} 
\end{aligned}  
\end{equation}  
where $\rho_0$ and $\eta_0$ are the initial states of system and environment, respectively, $\tU$ is a transformation representing all the system-environment interaction, and $e$ is a some effect.  By defining the state $\rho_{\rA\rE}  :  =  \tU (\rho_0\otimes \eta_0)$ the circuit of Eq. (\ref{simulation}) can be simplified to 
\begin{align}\label{simulation1}
   \begin{aligned}
    \Qcircuit @C=1em @R=.7em @! R {\multiprepareC{1}{\rho_{\rA\rE}}&\poloFantasmaCn{\rA}\qw&\qw\\
      \pureghost{\rho_{\rA\rE}}&\poloFantasmaCn{\rE}\qw&\measureD{e}}
  \end{aligned} \quad
   \begin{aligned}    
    \Qcircuit @C=1em @R=.7em @! R {=  & \quad &  \prepareC{\rho}&\poloFantasmaCn{\rA}\qw& \qw  &    \qquad   &  &\, .\\
    &&}
  \end{aligned}
  \end{align}
  
To capture the idea that the environment is discarded, we require the effect $e$ to be \emph{deterministic}:  
\begin{definition}
A \emph{simulation} of the preparation $\rho_\rA$  is a triple $( \rE,    \rho_{\rA\rE},  e )$ where $\rE$ is a system, $\rho_{\rA\rE}$ is a state of $\rA\otimes \rE$, and $e$ is a deterministic effect satisfying Eq. (\ref{simulation1}).    If the state $\rho_{\rA\rE}$ is pure, we say that $( \rE,    \rho_{\rA\rE},  e )$ is a \emph{pure simulation}---or, more concisely, a \emph{purification}---of $\rho_\rA$.  
  \end{definition}

Purifications arise, for example,  when we start from a \emph{pure} product state $\alpha_0\otimes \eta_0  \in  \Pur\St (\rA\otimes \rE)$ and evolve it through a \emph{reversible} transformation  $\tU$.      A purification gives  the agent  maximal control over the process of preparation: indeed, an agent possessing systems $\rA$ and $\rE$ can be sure that no side information can  hide  outside her laboratory.  

Given the importance of purifications, it is important to ask how many of them can be found for a given state. 
From a purification there are two trivial ways to generate  new ones: 
\begin{enumerate}
\item by transforming the environment with a reversible transformation   $\tU_\rE$  such that $(e| \, \tU  =  (e|$, and
\item by appending a dummy system $\rD$ to the environment, prepared in a pure deterministic state $\delta_\rD$  such that $\rho_{\rA\rE} \otimes \delta_\rD$ is pure.   
\end{enumerate}
We say that a pure simulation is \emph{essentially unique} if it is unique up to  trivial transformations: 
\begin{definition}
A state $\rho_\rA$ has an \emph{essentially unique purification} iff for every two purifications  $(\rE,  \Psi, e )$ and $(\rE', \Psi', e')$ with $\rE  =  \rE'$ one has  
\begin{align}\label{uniqueness}
 \begin{aligned}
    \Qcircuit @C=1em @R=.7em @! R {\multiprepareC{1}{\Psi'_{\rA\rE}}&\poloFantasmaCn{\rA}\qw&\qw\\
      \pureghost{\Psi'_{\rA\rE}}&\poloFantasmaCn{\rE}\qw&\qw}
  \end{aligned} 
  \quad     =  \quad
   \begin{aligned}
    \Qcircuit @C=1em @R=.7em @! R {   \multiprepareC{1}{\Psi_{\rA\rE}}&\poloFantasmaCn{\rA}\qw&\qw&\qw &\qw \\
      \pureghost{\Psi_{\rA\rE}}&\poloFantasmaCn{\rE}\qw&  \gate{\tU_\rE}  &  \poloFantasmaCn{\rE}\qw  &\qw}
  \end{aligned}   
\end{align}
and \footnote{It turns out that the second condition is automatically satisfied if the theory satisfies the Causality axiom---see the next section. }
\begin{align}
 \begin{aligned}
\Qcircuit @C=1em @R=.7em @! R {
    &\poloFantasmaCn{\rE} \qw &  \gate{{\tU_\rE}}  &\poloFantasmaCn{\rE} \qw   & \measureD{e'}      &  \qquad    &=  & \qquad &    \poloFantasmaCn{\rE} \qw   & \measureD{e}          }
\end{aligned}  ~. 
\end{align}  
for some reversible transformation $\tU_\rE$.     \end{definition}

\section{The principles}\label{sec:ax}

We are now ready to state our principles for quantum theory. We divide  them into five \emph{Axioms}  and one \emph{Postulate} \footnote{We differentiate the names in order to highlight the different roles of these principles in our reconstruction. Mathematically, there is no difference between axioms, postulates, background assumptions, and requirements in the OPT framework (all of them are ``axioms"). The point of using different names is just to provide a more intuitive picture.  
}.  
The five axioms are
\begin{enumerate}
\item[A1]{\bf Causality.}  No signal  can be sent from the future to the past.
\item[A2]{\bf Purity of Composition.} 
No side information can hide in the composition of two pure transformations. 
\item[A3]{\bf Local Tomography.}  State tomography can be performed with only local measurements. 
\item[A4]{\bf Perfect State Discrimination.} Every  normalized non-internal state  can be perfectly distinguished from some other
  state.
\item[A5]{\bf Ideal Compression.} Every  state   can be compressed in an ideal way. 
\end{enumerate}
The five Axioms express generic and rather unsurprising features, which are common to classical and quantum theory.  We regard the theories satisfying these axioms as \emph{standard}.  
The Postulate is
\begin{enumerate}
\item[P6]{\bf Purification.} Every preparation can be simulated via a pure preparation in an essentially unique way. 
% of a pure bipartite state of the system together with its environment. 
\end{enumerate}
Purification brings in a radically non-classical feature: the idea that randomness can be simulated    through the preparation of pure states.    We will see that this feature singles out quantum theory uniquely among all standard OPTs.

\subsection{Causality}

Causality states that signals cannot be sent from the future to the past.  To check this condition, it is sufficient to look at a special class of circuits, consisting  of a single preparation-test, followed by a single  observation-test.  Precisely, we have  the following 
\begin{prop}
An OPT satisfies Causality  if and only if for every  system $\rA\in\Sys$, every preparation-test $\boldsymbol{\rho}  \in
\Test  (\rI\to \rA,  \rX)  $ and every pair of  observation-tests ${\bf m_0} \in  \Test  (\rA\to \rI,  \rY_0) $ and ${\bf m_1} \in \Test  (\rA\to \rI,  \rY_1) $ one has 
   \[p(x|  {\bs m_0})    =    p(x|  {\bs m_1})  \qquad \forall x \in\rX \, ,  
   \]   
   with  
$ p(x|{\bs m_i})  :  =   \sum_{y_i\in\rY_i}   \, (  m_{y_i}|  \rho_x)$. 
\end{prop}
An even simpler condition for causality is given by \begin{prop}
  A theory satisfies Causality if and only if  every system $\rA$ has a unique deterministic effect $e_\rA \in\Det\Eff(\rA)$.
  \label{th:uniqdet}
\end{prop}

In categorical terms, the uniqueness of the deterministic effect can be phrased as  ``terminality of the tensor unit" in the category of deterministic transformations  $\Det\Transf$.   Categories where the tensor unit is terminal  have been introduced by Coecke and Lal    \cite{coecke2013causal, coecke2014terminality}, who named them   \emph{causal categories}.

Recall that deterministic effects can be used to describe   ``discarding operations", whereby a physical system is eliminated from the description.   Now, Causality is equivalent to the statement that every physical system can be discarded in a unique way.   Thanks to Causality, we can define the marginals of a bipartite state   in a canonical way 
\begin{definition}
Let $\rho_{\rA\rB}$ be a state of system $\rA\otimes \rB$.  The \emph{marginal} of $\rho_{\rA\rB}$ on system $\rA$  is the state $\rho_\rA$ defined as  
\begin{align*}
    \begin{aligned}    
    \Qcircuit @C=1em @R=.7em @! R {&  \prepareC{\rho_\rA}&\poloFantasmaCn{\rA}\qw& \qw     &\qquad&  :  =    }   
     \qquad \Qcircuit @C=1em @R=.7em @! R {\multiprepareC{1}{\rho_{\rA\rB}}&\poloFantasmaCn{\rA}\qw&\qw\\
      \pureghost{\rho_{\rA\rE}}&\poloFantasmaCn{\rB}\qw&\measureD{e}}
  \end{aligned}   
   \end{align*}
\end{definition}

\subsubsection{Causality and No-Signalling}
  An important  consequence of Causality is the impossibility to signal without interaction: in the lack of any interaction between system $\rA$ an system $\rB$, it is impossible to influence the
probability distribution of a test on system
$\rA$ by performing tests on system $\rB$.  The precise statement is the following 
\begin{prop}
For every state $\rho_{\rA\rB}$ and every triple of  tests $\bs \tA   \in \Test  (\rA\to \rA',  \rX)$,  $\bs \tB_0  \in  \Test (\rB\to \rB_0',  \rY_0)$ and $\bs  \tB_1  \in  \Test (\rB\to \rB'_1, \rY_1  )$ one has  
 \[p  \left(x|  {\bs \tB_0} \right)    =    p\left(x|  {\bs \tB_1}\right)  \qquad \forall x \in\rX \, ,  
   \]   
   with  
$ p(x|{\bs \tB_i})  :  =   \sum_{y_i\in\rY_i}   \, (  e_{\rB_i}|  \,  \tA_x \otimes \tB_{i, y_i}  \,  |     \rho_x)$,  $i\in\{0,1\}$. 
\end{prop} 

\medskip 

\subsubsection{Causality and conditional tests}
We introduced Causality as a negative statement: 
\begin{enumerate}
\item[{$\bf C$}:]  the choice of tests performed in the future \emph{cannot}  affect the outcome probabilities of  tests performed in the past.
\end{enumerate}
The axiom can  be reformulated in a positive, and slightly stronger way:  
\begin{enumerate}
\item[$\bf C' $:] the outcomes of tests performed in the past \emph{can} affect the choice of tests performed in the future.    
\end{enumerate}

Technically, Condition ${\bf C' }$  establishes the possibility of performing \emph{conditional tests}, defined as  follows: 
\begin{definition} 
 Given a test $\boldsymbol{\tT}  \in\Test(\rA\to \rB,  \rX)$   and a collection of tests $     \{       \bs \tS_x  \in  \Test (\rB\to \rC,  \rY_x)  ~|~  x\in \rX \}$,  the \emph{conditional test}  associated to them is the collection of transformations 
 \begin{equation*}
   \{  \bs \tS_x\} \odot \bs \tT   :  =   \left \{ \left. \Qcircuit @C=1em @R=.7em @! R {
&\qw \poloFantasmaCn{A}  &   \gate{\tT_x}  & \qw \poloFantasmaCn{B}   &  \gate{\tS_{y_x}^{x}}   & \qw\poloFantasmaCn{C}  &\qw}   \quad \right |\quad   x  \in  \rX  \,  ,  y_x  \in \rY_x    \right\}\, .
 \end{equation*}
 \end{definition}
 Condition $\bs \rC'$ states that such collection  is actually a \emph{test}, meaning that    
 \begin{enumerate}
 \item the set $    \rZ  =  \bigcup_{x\in\rX}   \, \{  x\}   \times \rY_x$ belongs to $\Outcomes$, and
 \item  the collection $  \{   \bs \tS_x  \}  \odot  \bs \tT  $ belongs to $\Test(\rA\to \rC, \rZ)$.    
 \end{enumerate}

The relation between  $\bs \rC$ and $\bs \rC'$ is the following: 
\begin{enumerate}
\item   $\bs \rC'$   \emph{implies} $\bs \rC$, 
\item  $\bs \rC$ implies that the theory can be enlarged  to another theory satisfying  $\bs \rC'$: thanks to $\bf C$, all conditional tests can be included without losing the consistency of the probabilistic structure \cite{puri}.
\end{enumerate} 
Since conditional tests can be included, we will always assume that they \emph{are} included,~i.~e.~we will take the validity of $\bs \rC'$ as part of the Causality package. 

\subsubsection{Convexity}  

 The ability to perform conditional tests brings naturally to  convexity of the sets of physical transformations.  This result can be obtained in two steps: 
 \begin{enumerate}
 \item   Under the standing assumptions that the theory is not deterministic and that the set $  \Transf (\rI\to \rI)$ is closed, we obtain that $  \Transf (\rI \to \rI)$ is  the whole  interval $[0,1]$. In other words, every number in the interval $[0,1]$ can be interpreted as the probability of some outcome in some test allowed by the theory.  
 \item Given two  transformations $\tT_0  ,  \tT_1  \in \Transf(\rA\to \rB)$, the convex combination $p \, \tT  +  (1-p)\,  \tT'$ can be generated by
 \begin{enumerate}
\item performing a binary test with the outcomes 0 and $1$ generated with probabilities $p_0= p$ and $p_1=1-p$  
\item conditionally on the occurrence of the  outcome   $i$, performing a test $\bs \tT_i$ containing the transformation $\tT_i$
\item coarse-graining over the appropriate outcomes of the conditional test. 
\end{enumerate}
\end{enumerate}

The above  observations show that convexity needs not be assumed from the start, but can be derived from non-determinism and Causality (in the  positive formulation $\bf C'$), under the standard assumption that the set of probabilities generated by tests in the theory is closed.

\subsubsection{Rescaling}  

In addition to convexity, conditional tests guarantee that every state is proportional to a \emph{normalized} state.  Specifically,   given a state $\rho$ of a generic system $\rA$, one can define the normalized state $\widetilde \rho    :  =   \rho/  (e_\rA|\rho)$.   An approximate way to prepare the state $\widetilde \rho$ is to 
\begin{enumerate}
\item pick a binary test $\{ \rho_0,\rho_1\}$ such that $\rho_1  =  \rho$ 
\item perform it $N$ times, generating a string of outcomes $(x_1,x_2,\dots, x_N)$  
\item perform a conditional test that  discards $N-1$ systems, keeping only a system $i$ such that $x_i  =1$, if such a system exists, or otherwise keeping only the first system
\item coarse-grain over all outcomes, thus obtaining the deterministic state   
\[ \rho_N :  =   (1-p_N) \,   \widetilde \rho   +  p_N  \,  \widetilde \rho_0     \qquad  \qquad p_N  =   (e_\rA  |\rho_0)^N \, . \]
\end{enumerate}
Clearly,  the state $\rho_N$  converges to $\widetilde \rho$ when $N$ goes to infinity.   Hence, the standard assumption that  the set of states is closed guarantees that $\widetilde \rho$ is a state allowed by the theory.

\subsection{Purity of Composition}

Purity of Composition is  a very primitive  rule about  how information propagates in time.   Mathematically, the axiom consists of the implication 
\begin{align*}
&\tA  \in  \Pur \Transf (\rA\to \rB)   \,  , \,   \tB  \in  \Pur\Transf(\rB\to \rC)   \\
&\qquad \qquad\qquad \qquad \qquad \Longrightarrow  \qquad  \tB\circ \tA   \in \Pur\Transf(\rA\to \rC) \, ,    
\end{align*}
required to be valid for all systems $\rA,\rB,\rC\in\Sys$ and for all pure transformations $\tA$ and $\tB$.

 Think of a world where this were not the case.  In that world, an  agent Alice could  perform a test   $\bs \tA   \in  \Test(\rA\to \rB \, , \rX)$ with such degree of control that, upon knowing the outcome,  she could not possibly know better what happened to her system.   Immediately after, another agent Bob could  perform another test $\bs \tB  \in  \Test (\rB\to \rC  \, , \rY)$ also having maximal knowledge of the system's conditional evolution.  Still, some of the resulting transformations  $\tB_y  \tA_x$  may not be pure.   This means that $\tB_y\tA_x$   can be simulated by a third party---Charlie---by performing one test $\{  \tC_z \}_{z\in\rZ}$ and joining together the outcomes in a suitable subset $\rS_{xy}  \subset \rZ$
 \begin{equation}\label{Charlieref}
\begin{aligned}
  \Qcircuit @C=1em @R=.7em @! R {
    &  \poloFantasmaCn {\rA} \qw& \gate{\tA_x}&\poloFantasmaCn{\rA}\qw&   \gate{  \tB_y}  &\poloFantasmaCn{\rA}\qw&  \qw
 }
 \end{aligned}
=
  \sum_{z\in\rS_{xy}}\begin{aligned}
  \Qcircuit @C=1em @R=.7em @! R {
    &  \poloFantasmaCn{\rA} \qw& \gate{\tC_z}&\poloFantasmaCn{\rA}\qw  &\qw}
 \end{aligned}  \, .
\end{equation}
Although this scenario is logically conceivable, it  rises a puzzling question:   What is the extra information about?    Which physical parameters correspond to the outcome $z$?   Surely the information is not about what happened in the first step, because Alice already had maximal knowledge about this.  Nor  it is about what happened in the second step, because Bob has maximal information about that.   The outcome $z$ has to specify  a feature of how the two time steps interacted together---in a sense, a kind of information that is  \emph{non-local in time}.         Quantum theory is non-local, but not in such an extreme way!  Indeed, pure transformations  in quantum theory are described by completely positive maps with a single Kraus operator,~i.~e.~of the form $\tA_x  (\cdot)   =  A_x \cdot A_x^\dag$ and $   \tB_y  (\cdot)   =   B_y \cdot  B_y^\dag$, and clearly the composition of two pure transformations is still  pure: $\tB_y  \tA_x (\cdot)  =   (B_yA_x)  \cdot   (B_yA_x)^\dag$.     Purity of Composition guarantees this property at the level of first principles.   

\subsection{Local  Tomography}

Local Tomography implies that even if a state is
entangled, the information it contains can be extracted by local
measurements. 
This fact  reconciles the holism of entanglement and the
reductionist idea that  the full information about a composite system can be obtained by studying its parts  \cite{maurobook}. 

Mathematically, Local Tomography states that  product effects form a separating set for the vector space $\St_\R  (\rA\otimes \rB)$.  Equivalently\footnote{Recall that we are assuming that the state spaces are finite-dimensional.}, they form a spanning set for the dual space $\St_\R(\rA\otimes \rB)^*  \equiv \Effr(\rA\otimes\rB)$.  
Hence, we must have the conditions  
\begin{align}\label{tensorproduct}
\Effr(\rA\otimes\rB)     =   \Eff_\R  (\rA)  \,  \otimes  \,  \Eff_\R  (\rB)   \qquad {\rm and} \qquad \St_\R(\rA\otimes\rB)     =   \St_\R  (\rA)  \,  \otimes  \,  \St_\R  (\rB)  \, ,
\end{align} 
 where $ \otimes$ in the r.h.s. denoted  the tensor product of finite dimensional vector spaces. 
 Eq. (\ref{tensorproduct}) implies that the dimensions of the vector spaces in question satisfy the product relation \cite{hardy01}
 \begin{align}
 D_{\rA\otimes \rB}      =    D_\rA  \,  D_\rB  \, .   
 \end{align} 

Moreover,  a generic state $\rho  \in  \St  (\rA  \otimes \rB)$ and a generic effect  $m  \in  \Eff  (\rA\otimes \rB)$ can be expanded as 
\begin{equation}\label{LTexpansion}
  \rho=\sum_{i,j}  \,   \rho_{ij}  \,  \left (v_i     \otimes   w_j  \right)    \qquad   {\rm and}  \qquad m=\sum_{i,j}  \,   m_{ij}\,     \left(v_i^*  \otimes w_j^*\right)  \, ,
\end{equation}
where $ [\rho_{ij}] $  and $ [m_{ij}]$  are real matrices,   $\{v_i\}_{i=1}^{D_\rA}$ and $\{w_j\}_{j=1}^{D_\rB}$  are bases  for the vector spaces $\St_\R(\rA)$ and $\St_\R(\rB)$, respectively, and $\{v^*_i\}_{i=1}^{D_\rA}$ and $\{w^*_j\}_{j=1}^{D_\rB}$  are the dual bases, defined by the relations  $(v_i^*|v_k)  =  \delta_{ik}$ and $(w_j^*|w_l)  =  \delta_{jl}$, respectively.   
As a result, the probability of the effect $m$ on the state $\rho$ can be expressed as 
\begin{align}\label{matrixprob}
(m|\rho)   =  \Tr  [   m  \,    \rho ]  \, ,
\end{align}
having committed a little abuse of notation in using the letter $m$ (respectively, $\rho$) both for the effect (respectively, state) and for the corresponding matrix $[m_{ij}]$  (respectively, $[\rho_{ij}]$).   
%This  equation will play an important role in our reconstruction of quantum theory.  

Finally, the decomposition  in Eq. (\ref{LTexpansion}) implies  the following 
\begin{theorem}
In a theory satisfying Local Tomography, the correspondence between a transformation  $ \tE  \in \Transf(\rA\to \rB)$ and the linear map 
  $\widehat \tE:\Str(\rA)\to\Str(\rB)$ is invertible.
\end{theorem}
\iffalse This result further highlights the importance of local
discriminability, because it is the principle that allows one to
characterize a transformation by local tests, without the need of
checking it on every type of bipartite system. In a sense, local
discrimination is the ingredient that makes the characterization of
transformations practically meaningful.
\fi
In other words, Local Tomography guarantees that physical transformations can be characterized in the simplest possible way: by preparing a set of input states and performing a set of measurements on the output.

A remarkable example of a theory that does not satisfy Local Tomography is quantum theory on real Hilbert spaces  \cite{stueckelberg1960quantum}, RQT for short.   In this theory,   states and effects  are real  symmetric  matrices,  and transformations are represented by  completely positive maps  mapping symmetric matrices into symmetric matrices.   The failure of the relation $D_{\rA\otimes \rB}  =  D_\rA  \, D_\rB$ was first noted by Araki  \cite{araki1980characterization}. More explicitly, Wootters \cite{wootters1990local} noted that two different bipartite states can be locally indistinguishable, as in the following extreme example: 
\begin{align}\label{rhorhoprime}
\rho  =   \frac 12   |\Phi_+\>\<\Phi_+| \,  + \,  \frac 12   |\Psi_-\>\<\Psi_-|  \,  \qquad  \qquad \rho'  =   \frac 12   |\Phi_-\>\<\Phi_-| \,  + \,  \frac 12   |\Psi_+\>\<\Psi_+| 
\end{align}
with $|\Phi_{\pm}\> : =   (   |0\>|0\>  \pm  |1\>|1\>  )/\sqrt 2$ and $|\Psi_{\pm}\>  =   (   |0\>|1\>  \pm  |1\>|0\>  )/\sqrt 2$.   Here  the  states $\rho$ and $\rho'$  have orthogonal support and therefore are perfectly distinguishable.  However, it is easy to check that one has  
\[  \rho  -  \rho'   =   \frac   12    \begin{pmatrix}    0  &  -1  \\  1  &  0  \end{pmatrix}  \otimes  \begin{pmatrix}    0  &  -1  \\  1  &  0  \end{pmatrix}  \, ,\]
 and, therefore, $  \Tr  [  (\rho- \rho')   (P_{\rA}\otimes P_{\rB})]  =  0$ for every pair of real symmetric matrices   $P_{\rA}$ and $P_{\rB}$. In other words, $\rho$ and $\rho'$ give exactly the same statistics for all possible local measurements.

RQT has  another, closely related quirk:  two \emph{different} transformations of system $\rA$ can act in the same way on all states of $\rA$. For example,  consider the qubit channels  $\tC$ and $\tC'$, whose action on a generic $2\times 2$ matrix is  defined by 
\begin{align*}
\tC  (  M) :  =  \frac 12    \, M   +  \frac 12 \,    Y M  Y   \qquad {\rm and} \qquad \tC' (M)  :  =  \frac 12   \, Z M Z   +  \frac 12   \,  X  M X     \, , 
\end{align*}
$X,Y,$ and $Z$ being the Pauli matrices.   When acting on symmetric  matrices, the two channels give exactly the same output:  one has $\tC  (\tau)  =  \tC'(\tau)  =  I/2$ for every symmetric matrix  $\tau $. 
On the other hand, one has  
\[   (\tC  \otimes \tI)   (  |\Phi_+\>\<\Phi_+)  =  \rho    \qquad (\tC'  \otimes \tI)   (  |\Phi_+\>\<\Phi_+)  =  \rho'   \, ,\]
where  $\rho$ and $\rho'$ are the two perfectly distinguishable states defined in Eq. (\ref{rhorhoprime}) above.   This means that, in fact, the two transformations $\tC$ and $\tC'$ are perfectly distinguishable with the help of a reference system.  For a more extensive discussion of tomography in  RQT we refer the reader to subsection V.A of Ref. \cite{puri} and to the work of Hardy and Wootters \cite{hardy2012limited}.

\subsection{Perfect State Discrimination}

Perfect State Discrimination is an optimistic statement about the possibility to encode bits without error.   It guarantees that every state that \emph{could}  be part of a set of perfectly distinguishable states \emph{is} indeed perfectly distinguishable from some other state. 

By virtue of Perfect State Discrimination,  every normalized non-internal state $\rho_0 $  can be perfectly distinguished from some state $\rho_1$.   As a result, the two states $\rho_0$ and $\rho_1$ can be used to encode the value of a bit without errors.  It is easy to see that Quantum theory satisfies the axiom. Indeed,  Aadensity matrix is internal if and only if it has full rank. Hence, a non-internal density matrix $\rho_0$ must have a kernel, so that  every state $\rho_1$ with support in the kernel of $\rho_0$ will be perfectly distinguishable from $\rho_0$.   
%The situation is completely analogue in classical probability theory: a probability distribution   $p_0(x)$ over some finite set $\rX$  is non-internal if and only if its support is not the whole set $\rX$, and, therefore, every probability distribution  $p_1(x)$ with support in the complement of the support of $p_0(x)$ will be perfectly distinguishable from it.        

\subsection{Ideal Compression}

Ideal Compression   expresses the idea that  information is \emph{fungible},~i.~e.~independent of the physical support in which it is encoded.   
The axiom  implies non-trivial statements about the state spaces arising in the theory. For example, suppose that  the theory contains a system whose space of deterministic states is a square.  Then,  the theory  should  contain  also a system whose space of deterministic states is a segment---in other words, the theory  should contain a classical bit. Indeed, only in this way one could encode a side of the square in a lossless and maximally efficient way.    More generally,  Ideal Compression imposes that the every face of the convex set of deterministic states  be in one-to-one correspondence with  the set of deterministic states of some smaller physical system.
  
Ideal Compression is clearly satisfied by quantum theory. Indeed,  every density matrix of rank $r$  can be compressed ideally to a density matrix of an $r$-dimensional quantum system.  For example, the two-qubit density matrix   
 \begin{align}
 \rho  =     \begin{pmatrix}
 \rho_{00,00}  &    0  &  0   &    \rho_{00,11}  \\
 0&0&0&0\\
 0&0&0&0\\
 \rho_{11,00}&0&0&\rho_{11,11}
  \end{pmatrix}
 \end{align}  
can be compressed ideally to the one-qubit density matrix  
\begin{align}
\tE  (\rho)   =  \begin{pmatrix}  \rho_{00,00}  &  \rho_{00,11}  \\
\rho_{11,00}  &  \rho_{11,11}  
\end{pmatrix}      
\end{align}  
with  encoding and decoding channels given by  
\begin{align*}
\tE  (\cdot)  &:  =    V^\dag  \, (\cdot) \, V   +   \Tr  [  (  I  -  V\,V^\dag)  \,   (\cdot) \, ] \,  |0\>\<0|  \qquad   \qquad V :  =|0\>|0\>  \<0| + |1\>|1\>  \<1|\\
\tD  (\cdot)    &:  =   V  \, (\cdot) \,  V^\dag  
 \, . \end{align*}

Note that Ideal Compression refers to a \emph{single-shot, zero error scenario},~i.~e.~a scenario where the source is used only once and no decoding errors are allowed.   
  Such a scenario is different from the asymptotic scenario considered in Shannon's \cite{shannon}  and Schumacher's \cite{schumacher1995quantum} compression, wherein small decoding errors are allowed, under the condition that they vanish in the asymptotic limit of infinitely many  uses of the same source.

\subsection{Purification}

While our first five axioms expressed standard requirements for information-processing, Purification brings in a radically new idea:  at least in principle, every state can be prepared by an agent who has maximal control over all the systems involved in the preparation process.  
In short, Purification allows  us to  harness randomness by controlling the environment.  The idea does not apply only to preparations, but also to arbitrary deterministic transformations:  combining Purification with Causality and Local Tomography, we can prove the following  

\begin{theorem}{\cite{puri}}\label{th:purrevdil}
  For  every deterministic transformation  $\tT  \in \Det\Transf(\rA\to \rA')$, there
  exist two systems $\rE$ and $\rE'$, a pure state $\eta \in\Pur\St(\rE)$, and a
  reversible  transformation  $\tU  \in \Rev\Transf  (\rA\otimes \rE  \to  \rA'\otimes \rE')$ such that
  \begin{equation}
    %\begin{aligned}
    %  \Qcircuit @C=1em @R=.7em @! R {
      %  &\poloFantasmaCn{\rA}\qw&\gate{\tT}&\poloFantasmaCn{\rA'}\qw&\qw}
    %\end{aligned} 
   % =
    \begin{aligned}
      \Qcircuit @C=1em @R=.7em @! R {
      &\poloFantasmaCn{\rA}\qw&\gate{\tT}&\poloFantasmaCn{\rA'}\qw&\qw    &    =     & &\poloFantasmaCn{\rA}\qw&   \multigate{1}{\tU}  &          \poloFantasmaCn{\rA'}\qw    &    \qw  \\
       &&&&&  &\prepareC{\eta} &\poloFantasmaCn{\rE}\qw&  \ghost{\tU}  &          \poloFantasmaCn{\rE'}\qw    &    \measureD{e}   }
    \end{aligned} \, ,
    \label{eq:manyw?}
  \end{equation} 
  where $e$ is the unique deterministic effect of system $\rE'$. 
\end{theorem}
In other words, Purification implies that every irreversible process can be simulated through  reversible interactions between the system and its environment, with the environment initialized in a pure state.   This result is a necessary condition for the formulation of physical theories in which elementary processes are reversible at the fundamental level.

Purification is known to be satisfied by quantum mechanics.  For example, consider a  single-qubit mixed state, diagonalized as 
\begin{equation}
  \rho=p_0|0\>\<0|+p_1|1\>\<1| \, ,
\end{equation}
for some suitable orthonormal basis $\{|0\>, |1\>\}$.  
A purification of the state $\rho$ can be obtained by adding a second  qubit  and by preparing the two qubits in the pure state 
\begin{equation}
  |\Psi\>:=\sqrt{p_0} \, |0\> |0\> +\sqrt{p_1}\,  |1\> |1\> \, .
\end{equation}
Indeed, it is immediate to see that $\rho$ is the marginal of the density matrix $|\Psi\>\<\Psi|$  on the first qubit.     In addition, any other purification  $|\Psi'\> $---using a single qubit as the purifying system---must be of the form $|\Psi\>'   =   (I \otimes U)  \,|\Psi\>$ for some unitary matrix $U$.   

In the quantum information community, taking purifications is a standard approach to quantum communication, cryptography, and quantum error correction.      The approach is    familiarly known with the nickname of ``going to the Church of the larger Hilbert space" \footnote{The  expression is due to John Smolin, see e.g. the lecture notes \cite{bennett}.}.     
Purification is known among mathematicians as the \emph{Gelfand-Naimark-Segal construction} \cite{GN,S}.  

Two important remarks are in order:
 \begin{enumerate}
 \item {\em Purification, entanglement, and quantum information.}   
 Purification is intimately linked with the phenomenon of \emph{entanglement}  \cite{schrodinger1935discussion}, namely the existence of pure bipartite states   $\Psi_{\rA\rB} $  that are not of the product form  $\psi_\rA  \otimes \psi_\rB$.  
 In the OPT framework, the link is made precise by the following 
 \begin{prop}
 Let $\Theta$ be a theory satisfying Causality, Local Tomography, and Purification.  Then, there are only two alternatives:  either $\Theta$ is deterministic, or $\Theta$ exhibits entanglement.  
\end{prop}
Under our standing assumption that the theory is non-deterministic, entanglement follows from Purification as a necessary consequence.  

Entanglement is a very peculiar feature---far from what  we experience in our
everyday life.  How can we claim that we know $\rA$ \emph{and} $\rB$ if we do not know $\rA$ alone?
This puzzling feature had been noted already in the early days of quantum theory, when 
Schr\"odinger famously wrote: ``Another way of expressing the peculiar situation is: \emph{the best
  possible knowledge of a whole does not necessarily include the best possible knowledge of all its
  parts}"  \cite{schrodinger1935discussion}. And, in the same paper: ``I would not call that \emph{one} but rather
\emph{the} characteristic trait of quantum mechanics, the one that enforces its entire departure
from classical lines of thought".    
In a sense, our reconstruction can be considered as a mathematical proof of Schr\"odinger's intuition \footnote{It is worth stressing that Schr\"odinger's paper was not just about the \emph{existence} of entangled states, but also about how entanglement interacted with the reversible dynamics and with the process of measurement (cf. the notion of \emph{steering}, which made its first appearance in the very same paper).}:  on the background of  five standard axioms satisfied by both classical theory and quantum theory,  Purification is the ingredient that allows to reconstruct the Hilbert space framework and  the distinctive information-theoretic features of quantum theory.    Combined with Causality and Local
Tomography, Purification  already reproduces an impressive list of quantum-like features,  like no-cloning, no-programming,
information-disturbance tradeoff, no bit commitment, conclusive teleportation and entanglement swapping, the reversible dilation of  channels,
the state-transformation isomorphism, the structure of error correction, and the structure of no-signalling channels \cite{puri}.

\item {\em Purification and the Many Wolrd Interpretation. }   Pondering about the meaning of  Purification, 
one may tempted to conclude  that it  favours the
Many Worlds Interpretation (MWI) of quantum mechanics \cite{everett1957relative}.   In fact,   Purification is feature of quantum theory, and, as such, it does not favour the MWI  more than quantum theory itself does.  Whether or not quantum theory provides any evidence for many worlds is a debatable point, but  the validity of Purification is independent such interpretative issue.    Furthermore, we stress that we did not phrase  Purification as  an ontological statement about ``how processes   occur in nature", but rather an operational  statement about the agent's ability  to simulate  physical processes with maximal control.  Purification is \emph{compatible} with the idea that processes are reversible at the fundamental level, and its validity is a \emph{necessary condition} for building up a physical description of nature in terms of pure states and reversible processes. Still, here we do not make any commitment as to how processes are realized in nature, because this would unnecessarily limit the range of application of our results.  

\end{enumerate}

\iffalse 
Interestingly, the converse is also true: the existence of minimal purifications \emph{implies} Ideal Compression.    To see that, let us formulate a stronger version of the Purification postulate:
\begin{postulate}[Minimal Purification]  
Every state  has a minimal purification.  If  two (non necessarily minimal) purifications of the same state have the same purifying systems, then they are equivalent up to a reversible transformation on the purifying system. 
\end{postulate}
Then, it is quite easy to show the following 
\begin{prop}
An OPT satisfies Minimal Purification if and only if it satisfies Purification and Ideal Compression.  
\end{prop}
\fi

\section{The reconstruction of Quantum Theory}\label{sec:reconstruction}

Here we provide  a summary of the reconstruction of  Refs. \cite{puri,deri},  highlighting the key theorems  and providing a  guide to the original papers. The scope of the  reconstruction is not just to derive the Hilbert space framework, but also to rebuild the key quantum features directly from first principles.  
Accordingly, we try to derive as much as possible of quantum theory directly from the axioms, leaving Hilbert spaces to the very end.  
We organize our results in six subsections:
\begin{enumerate}
\item Elementary facts.    
\item Correlation structures. 
\item Distinguishability structures.  
\item Interaction between correlation and distinguishability structures.  
\item Qubit features.
\item The density matrix. 
\end{enumerate}  
% It worth mentioning that the scope of these works was

% In terms of potential for future developments, one could even argue that reconstructing from basic principles the  connections among different quantum features  is even more promising than just getting to Hilbert spaces and operators. 

\subsection{Elementary facts}   
%We collect here some basic results that will be used often in the course of the reconstruction.

\subsubsection{From Local Tomography}  

Local Tomography implies a few  useful facts:
\begin{enumerate}
\item  If  $\alpha  \in \St(\rA)$ and $\beta\in\St(\rB)$ are pure, then also $\alpha\otimes \beta$ is pure. 
\item Let $\rho_{\rA \rB}$ be a state of the composite system $\rA\otimes \rB$ and, assuming Causality, let $\rho_\rA$ be its marginal on system $\rA$.  If $\rho_\rA$ is pure, then $\rho_{\rA\rB} $ is a product state. 
\item If $\rho_\rA\in\St(\rA)$ and $\rho_\rB\in\St(\rB)$ are internal states, then also $\rho_\rA\otimes \rho_\rB$ is an internal state.  
\item Suppose that every system $\rA$ has a unique \emph{invariant state} $\chi_\rA$,~i.~e.~a unique state satisfying the condition $\tU  \chi_\rA  =  \chi_\rA$ for every reversible transformation $\tU$.      Then,  $\chi_{\rA\otimes \rB}  =  \chi_\rA  \otimes \chi_\rB$.  
\end{enumerate}

\subsubsection{From Purification}  

Purification has a few immediate consequences.  First, all pure states of a given system are connected to one another through reversible transformations: 
\begin{prop}
For every system $\rA\in\Sys$ and every pair of pure states $\alpha,  \alpha'  \in  \Pur\St(\rA)$ there exists a reversible transformation  $\tU$ such that $\alpha'  =  \tU  \,\alpha$.
\end{prop}
To prove this fact, it is enough to pick a system $\rB$ and  pure state $\beta  \in \Pur\St(\rB)$, consider the states  $\Psi  = \alpha\otimes \beta$ and $\Psi'=\alpha'  \otimes \beta$ as purifications of  $\beta$, and invoke the essential uniqueness of purification [Eq. (\ref{uniqueness})].  Mathematically, the above proposition expresses the fact that the action of the reversible transformations is \emph{transitive} on the manifold of pure states---a requirement that played an important role in many recent reconstructions, see~e.~g.~\cite{hardy01,dakic11,masanes11}.  
A  byproduct of transitivity   is 
\begin{prop} Every system  $\rA\in\Sys$ has a unique invariant state $\chi_\rA$. 
\end{prop}

Finally, combining Ideal Compression and Purification it is easy to see  that every state  has a \emph{minimal purification}, in the following sense
\begin{definition}
Let $\Psi  \in  \Pur\St(\rA\otimes \rB)$ be a pure state with marginals $\rho_\rA$ and $\rho_\rB$ on systems $\rA$ and $\rB$, respectively.  We say that $\Psi$ is a \emph{minimal purification} of $\rho_\rA$ iff $\rho_\rB$ is internal.  
\end{definition}
To construct a minimal purification, it is enough to take an arbitrary purification and to compress the state of the purifying system.  
%Minimal purifications will play a crucial role in the following paragraphs. 
%Now, an important property of minimal purifications is that the purifying system is uniquely defined, up to operational equivalence.  

\subsection{Correlation structures}

%Here we focus on the way in which physical systems can be correlated and on the way such correlations can be exploited for information-processing tasks, such as entanglement swapping and teleportation.     

\subsubsection{Pure Steering}  

One of the most important consequences of our axioms is that pure bipartite states  enable  \emph{steering}, namely the ability to remotely generate every desired ensemble decomposition of a marginal state \cite{schrodinger1935discussion,barnum2013ensemble}:
\begin{prop}[Pure Steering]\label{prop:steering}
Let $\Psi$ be a pure state of the composite system $\rA\otimes \rB$,  let $\rho$ be the marginal of $\Psi_{\rA\rB}$ on system $\rA$, and let ${\bs \rho}  =  \{\rho_x\}_{x\in\rX}$ be an ensemble decomposition of $\rho$. Then there exists a measurement $\bs b  =  \{b_x\}_{x\in\rX}$ such that   
 \begin{align}
   \begin{aligned}
    \Qcircuit @C=1em @R=.7em @! R {\multiprepareC{1}{\Psi}&\poloFantasmaCn{\rA}\qw&\qw\\
      \pureghost{\Psi}&\poloFantasmaCn{\rB}\qw&\measureD{b_x}}
  \end{aligned} \quad
   \begin{aligned}    
    \Qcircuit @C=1em @R=.7em @! R {=  & \quad &  \prepareC{\rho_x}&\poloFantasmaCn{\rA}\qw& \qw  &    \qquad   &\pureghost{ \phantom{supercali}}& \forall x\in\rX  \, .\\
    &&}
  \end{aligned}
  \end{align}
\end{prop}

Pure steering is the essential ingredient for a number of major results. 
The first result is the existence of \emph{pure, tomographically faithful states}.    A state  $\rho  \in  \St(\rA\otimes \rB)$ is called \emph{tomographically faithful} for system $\rA$ iff the implication       
 \begin{align}
   \begin{aligned}
    \Qcircuit @C=1em @R=.7em @! R {\multiprepareC{1}{\rho}&\poloFantasmaCn{\rA}\qw&\gate{\tT}  &  \poloFantasmaCn{\rC}\qw  & \qw   \\
      \pureghost{\rho}&\poloFantasmaCn{\rB}\qw&  \qw & \qw &\qw  }
  \end{aligned} 
=  \quad    \begin{aligned}
    \Qcircuit @C=1em @R=.7em @! R {\multiprepareC{1}{\rho}&\poloFantasmaCn{\rA}\qw&\gate{\tT'}  &  \poloFantasmaCn{\rC}\qw  & \qw   \\
      \pureghost{\rho}&\poloFantasmaCn{\rB}\qw&  \qw & \qw &\qw  }
  \end{aligned}   \qquad \Longrightarrow  \qquad      \tT  =  \tT'  \, ,
\end{align}
holds for every system $\rC$ and every pair of transformations  $\tT$ and $\tT'$  of type $\rA\to \rC$.   Thanks to Pure Steering and Local Tomography, we are able to construct tomographically faithful states: 
\begin{prop}
Let $\rho_\rA  $ be an internal state of system $\rA$ and let $\Psi  \in \Pur\St(\rA\otimes \rB)$ be a purification of $\rho_\rA$. Then,   $\Psi$ is tomographically faithful for system $\rA$.  
\end{prop} 

The result can be  improved by choosing a \emph{minimal} purification: in this way, the pure state $\Psi$ is faithful on both systems $\rA$ and $\rB$.   We call such a state \emph{doubly faithful}. 
\subsubsection{Conjugate systems}
  A canonical choice of doubly faithful state is obtained by picking a minimal purification of  the invariant state $\chi_\rA$.   
  We denote such purification by $\Phi  \in  \Pur\St(\rA\otimes \over \rA)$ and call system $\overline \rA$ the \emph{conjugate of system $\rA$}.   
  The name is motivated by the following facts:  
  \begin{enumerate}
  \item system $\over \rA$ is uniquely defined, up to operational equivalence%\footnote{This fact follows from the essential uniqueness of purification, combined with the fact that the state $\Phi$ is doubly faithful.} 
  \item   the marginal of $\Phi$ on system $\over \rA$ is the invariant state $\chi_{\over \rA}$  (cf. Corollary 46 of \cite{puri}), meaning that we have $\over {\over  \rA }   =  \rA$, up to operational equivalence. 
  \end{enumerate} 
Summarizing, the state $\Phi$ satisfies the relations
   \begin{align}
   \begin{aligned}
    \Qcircuit @C=1em @R=.7em @! R {\multiprepareC{1}{\Phi}&\poloFantasmaCn{\rA}\qw&\qw\\
      \pureghost{\Phi}&\poloFantasmaCn{\over \rA}\qw&\measureD{e}}
  \end{aligned} \quad
   \begin{aligned}    
    \Qcircuit @C=1em @R=.7em @! R {=  & \quad &  \prepareC{\chi}&\poloFantasmaCn{\rA}\qw& \qw    \\
    &&}
  \end{aligned}  \qquad  {\rm and} \qquad \begin{aligned}
    \Qcircuit @C=1em @R=.7em @! R {\multiprepareC{1}{\Phi}&\poloFantasmaCn{\rA}\qw&\measureD{e}\\
      \pureghost{\Phi}&\poloFantasmaCn{\over \rA}\qw&\qw}
  \end{aligned} \quad
   \begin{aligned}    
    \Qcircuit @C=1em @R=.7em @! R {
    &&\\=  & \quad &  \prepareC{\chi}&\poloFantasmaCn{\over \rA}\qw& \qw    }   
  \end{aligned}~.
 \end{align}
By  analogy with 	quantum theory, we call  $\Phi$ a \emph{Bell state}.   

\subsubsection{The state-transformation isomorphism}

For a given transformation $\tT $, we define the (generally unnormalized) state   
\begin{align}
   \begin{aligned}
    \Qcircuit @C=1em @R=.7em @! R {\multiprepareC{1}{\Phi_{\,\tT}}&\poloFantasmaCn{\rC}\qw& \qw   \\
      \pureghost{\Phi_{\,\tT}}&\poloFantasmaCn{\over \rA}\qw&  \qw }
  \end{aligned}   \quad
  :=   \quad   \begin{aligned}
    \Qcircuit @C=1em @R=.7em @! R {\multiprepareC{1}{\Phi}&\poloFantasmaCn{\rA}\qw&\gate{\tT}  &  \poloFantasmaCn{\rC}\qw  & \qw   \\
      \pureghost{\Phi}&\poloFantasmaCn{\over \rA}\qw&  \qw & \qw &\qw}
  \end{aligned}   ~ .
  \end{align} 
 and call the correspondence $ \tT  \mapsto     \Phi_\tT $ the \emph{state-transformation isomorphism}.  Since the Bell state $\Phi$ is doubly faithful, the correspondence is one-to-one.  
 %The main properties of the state-transformation isomorphism are 
% \begin{enumerate}
% \item \emph{Injectivity:}   $ \Phi_\tT  =  \Pji_{\tT'} $ implies $$
 %\end{enumerate}
  In quantum theory, the state-transformation isomorphism  concides with  the Choi isomorphism \cite{choi1975completely}.  By analogy, we call the state $\Phi_\tT$ the \emph{Choi state}. 
  %, whose operational interpretation was given more than twenty years after the original paper \cite{debbieleung,maurochoi,cirac}.  

A powerful byproduct  of the state-transformation isomorphism is that the  normalized states completely identify the theory: 
\begin{theorem}
  Let $\Theta$ and $\Theta'$ be two theories with the same set of systems.   If the sets of normalized states of $\Theta$ and $\Theta'$ coincide for all systems,  then the two 
  theories coincide.
  \label{th:staspec}
\end{theorem}

Thanks to this result, deriving the density matrix representation of normalized states is sufficient to derive the whole of quantum theory.

\subsubsection{Conclusive entanglement swapping}
 An important consequence of Pure Steering  is the possibility of entanglement swapping, namely the possibility to generate entanglement remotely by performing a joint measurement.   
Consider, as a prototype of entangled state, the Bell state $\Phi $.  
 Then, it is possible to show that there exists a pure effect $E  \in\Pur\Eff(\over \rA\otimes \rA)$ and a non-zero probability $p_\rA>0$ such that 
\begin{equation}\label{entswap}
\begin{aligned}
\Qcircuit @C=1em @R=.7em @! R {\multiprepareC{1}{\Phi}& \qw \poloFantasmaCn {\rA_{\phantom{1}}} &\qw \\
\pureghost{\Phi} & \qw \poloFantasmaCn {\rB_1} &\multimeasureD{1} {E}  \\
\multiprepareC{1}{\Phi}& \qw \poloFantasmaCn {\rB_2} &\ghost{E} \\
\pureghost{\Phi} & \qw \poloFantasmaCn {\rC_{\phantom{1}}} &\qw}  
\end{aligned}  
~=~p_\rA ~  \begin{aligned} 
\Qcircuit @C=1em @R=.7em @! R 
{\multiprepareC{3}{\Phi}& \qw \poloFantasmaCn \rA &\qw   &&\\ 
&&    & \pureghost{\phantom{supercali}}  &   \rA   \equiv  \rB_2 \, ,\phantom{\equiv \over \rA}      \\
&&   &   \pureghost{\phantom{supercali}}  &   \rB_1  \equiv  \rC  \equiv \over \rA \, .    \\
\pureghost{\Phi}& \qw \poloFantasmaCn \rC &\qw &&    } \end{aligned}
\end{equation} 
This diagram represents an instance of  \emph{conclusive entanglement swapping}:  conditionally on the occurrence of the effect $E$, the two systems $\rA$ and $\rC$ are prepared in the Bell state, consuming the initial entanglement  present in the composite systems $\rA\otimes \rB_1$ and $\rB_2\otimes \rC$.    

The possibility of entanglement swapping follows easily from Pure Steering:  Since the states $\chi_\rA$ and $\chi_{\over \rA}$ are internal, Local Tomography implies that their product $\chi_\rA  \otimes \chi_{\over \rA}$  is internal.   Hence, there must exist a non-zero probability $p_\rA  > 0$ such that 
\begin{align}\label{probtele}  \chi_\rA  \otimes \chi_{\over \rA}    =    p_\rA  \,   \Phi   +    (1-p_\rA)  \,   \tau  \, , 
\end{align}
for some state $\tau$.     Applying Pure Steering  (proposition \ref{prop:steering}) to the pure state $\Phi\otimes \Phi$ and to the ensemble $ \{p_\rA \, \Phi\, ,  (1-p_\rA)  \,\tau\}$ one  can find a binary measurement  $  \{   E  ,   e_{\rB_1}  \otimes e_{\rB_2}  -  E \}$  such that the entanglement swapping condition  (\ref{entswap})  holds.  
  Using the fact that the state $\Phi\otimes \Phi$ is doubly faithful, it is  easy to see that  the effect $E$ must be pure.

\subsubsection{Conclusive teleportation}
By the state-transformation isomorphism,  conclusive entanglement swapping is equivalent to \emph{conclusive teleportation} \cite{telep}, expressed by the diagram    
\begin{equation}\label{tele}
\begin{aligned}
\Qcircuit @C=1em @R=.7em @! R {\multiprepareC{1}{\Phi}& \qw \poloFantasmaCn {\rA} &\qw \\
\pureghost{\Phi} & \qw \poloFantasmaCn {\over \rA} &\multimeasureD{1} {E}  \\
  & \qw \poloFantasmaCn {\rA} &\ghost{E} }  
\end{aligned}  
~=~p_\rA ~  \begin{aligned} \Qcircuit @C=1em @R=.7em @! R {&\qw & \qw \poloFantasmaCn \rA & \qw  &\qw} \end{aligned}   \quad .
\end{equation} 
Indeed,   the entanglement swapping  diagram  (\ref{entswap}) is equivalent to the condition  $  \Phi_{\tT}  =  \Phi_{\tT'}$, with 
\begin{equation}\label{ttprime}
\begin{aligned}
\Qcircuit @C=1em @R=.7em @! R {  & \qw \poloFantasmaCn {\rA} &   \gate{\tT}  &  \qw \poloFantasmaCn {\rA} &  \qw }
\end{aligned}  :  = \quad
\begin{aligned}
\Qcircuit @C=1em @R=.7em @! R {\multiprepareC{1}{\Phi}& \qw \poloFantasmaCn {\rA} &\qw \\
\pureghost{\Phi} & \qw \poloFantasmaCn {\over \rA} &\multimeasureD{1} {E}  \\
  & \qw \poloFantasmaCn {\rA} &\ghost{E} }  
\end{aligned}  \qquad{\rm and} \qquad  
\begin{aligned}
\Qcircuit @C=1em @R=.7em @! R {  & \qw \poloFantasmaCn {\rA} &   \gate{\tT'}  &  \qw \poloFantasmaCn {\rA} &  \qw }
\end{aligned}  :  = \quad  
 \begin{aligned} \Qcircuit @C=1em @R=.7em @! R {   & \qw \poloFantasmaCn \rA &  \gate{ p_\rA\,  \tI_\rA}  &    \qw \poloFantasmaCn \rA  &\qw} 
 \end{aligned}   
 \quad .
\end{equation}   
By the state-transformation isomorphism, $\Phi_\tT  =  \Phi_{\tT'}$ implies $\tT  = \tT'$, which is nothing but the teleportation diagram. 

\subsubsection{The teleportation upper bound}

Combined with Local Tomography, the teleportation diagram allows us to upper bound the dimension of the state space.  The idea is to write the teleportation diagram in matrix elements, by expanding $\Phi$ and $E$ as
\begin{align}\label{PhiE}
\Phi   =   \sum_{ik}   \, \Phi_{ik}  \,    \left(    v_i  \otimes  w_k   \right)    \qquad {\rm and}\qquad E  =  \sum_{jl} E_{jl}  \, \left(  w_j^*  \otimes v_l^* \right)  \, ,  
\end{align}
with suitable bases $\{v_i\}_{i=1}^{D_\rA}$  and $\{w_j\}_{j=1}^{D_{\overline \rA}}$.    In this representation, Eq. (\ref{tele})  becomes  
\begin{align}   
    [\,\Phi \,  E\,]_{il}   =   p_\rA  \,     \delta_{il}   \, ,
\end{align}
and, taking the trace,  
\begin{align}\label{trace} 
\Tr  [ \Phi  E]     =    p_\rA  \,  D_\rA  \, .
\end{align}
On the other hand, we have   
\begin{align}
\Tr  [ \Phi  E]     =   ( E |  \, \tS_{ \rA, \over \rA} \,   |    \Phi )   \le 1  \, ,
\end{align}
which combined with Eq. (\ref{trace})  leads  to bound  
\begin{align}\label{teleupper}
D_\rA  \le \frac 1{p_\rA}  \, .
\end{align}
Clearly, in order to have the best bound we need to find the \emph{maximum} probability of teleportation. To discover  what the maximum is, we need to move our attention to the distinguishability structures implied by our axioms.

\subsection{Distinguishability structures}
%We now  consider the task of distinguishing states, proving a number of results on state discrimination and on the structure of perfectly distinguishable states.  This part of the reconstruction focuses on single-system features, such as the spectral theorem, and is the part  that connects more closely with the tradition of quantum logic.   

\subsubsection{No disturbance without information}
Our first move is to derive a simple result about the structure of measurements: a measurement that extracts no information from a face of the state space can be implemented without disturbing that face.   By \emph{face of the state space} we mean a face of the convex set of deterministic states\footnote{We recall that a \emph{face} of a convex set $C$ is a convex subset $F\subseteq  C$ satisfying the condition that, for every $x  \in  F$, if $x$ is a non-trivial convex combination of $x_1$ and $x_2$ with $x_1,x_2 \in  C$, then $x_1$ and $x_2$ belong to $F$.}.   We say that the measurement $\bs m  \in \Test(\rA\to \rI,\rX) $ \emph{does not extract information}   from the face  $F$  iff  there exists a set of probabilities $\{p_x\}_{x\in\rX}$ such that
\begin{align*}
\left(  m_x |  \tau\right)  =    p_x  \qquad   \forall x\in\rX,  \quad \forall \tau\in F \, .
\end{align*}
Also, we say that a test $\bs  \tT  \in \Test (\rA\to\rA,  \rX)$ \emph{does not disturb the face $F$}  iff $\sum_{x\in \rX}   \tT_x   |\tau)  =  |\tau)$ for every state $\tau  \in  F$.    

With this terminology, our  result is the following:  
\begin{prop}
If a measurement $\bs m$ does not extract information from the face $F$, then there exists a test  $\bs \tT$ that realizes the measurement---namely    $(e_\rA | \tT_x  =  m_x$,  $ \forall x\in\rX$---and does not disturb $F$.
\end{prop} 

This result has two important consequences. First, it allows us to establish whether or not a set of perfectly distinguishable set can be extended:  
\begin{prop}\label{prop:maximal}
Let  $\rS  = \{\rho_x~|~x \in  \rX\}$ be a set of perfectly distinguishable states and let $\omega_\rS$ be its barycenter, defined as  
\[   \omega_\rS  :=   \frac 1 {|\rX|}  \,  \sum_{x\in\rX}    \,  \rho_x  \, . \]
Then, the following are equivalent:  
\begin{enumerate}
\item the set $\rS$ is \emph{maximal},~i.~e.~no other set   $\rS'  \supset  \rS$ can consist  of perfectly  distinguishable states
\item the barycenter of $\rS$ is internal. 
\end{enumerate}
\end{prop}

Another important consequence is that only  the \emph{pure} maximal sets can have maximum cardinality: 
\begin{prop}\label{prop:onlypure}
 Let  $\rS $ be a maximal set of perfectly distinguishable states of system $\rA$.  If one of the states in $\rS$ is not pure, then there exists another maximal set $\rS'  \subset \St(\rA)$, consisting only of pure states and having strictly larger cardinality $|\rS'|  >  |\rS|$.   
\end{prop}
Combining the above points we have that every pure state belongs to some maximal set of perfectly distinguishable pure states.   For short, we call such such sets \emph{pure maximal sets}.

 \subsubsection{Duality between pure states and pure effects}
For a pure maximal set  $\rS$, %  $\{  \alpha_n  ~|~  n=  1,\dots, N\} $, let  $\bs m  $ be the measurement  
we observe that the measurement that distinguishes the states in $\rS$ must consist of \emph{pure} effects.    Hence, for every pure state $\alpha \in  \Pur\St (\rA)$ there exists an pure effect $a$ such that $(a|\alpha)  =1$.   Expanding on this observation, we 
  establish a one-to-one correspondence between pure normalized states and pure normalized effects \footnote{ We call an  effect  of system $\rA$ \emph{normalized} iff there exists an effect a state $\rho$ such that $(a|\rho)  =1$.}, denoted by  $\Pur\St_1(\rA)$   and $\Pur\Eff_1  (\rA)$, respectively. 
  \begin{theorem}
For every system $\rA\in\Sys$, there exists a one-to-one map $\dag:  \Pur\St_1(\rA)\to \Pur\Eff_1  (\rA) $, sending pure normalized states to pure normalized effects and satisfying the condition  
\[  (  \alpha^\dag | \alpha  )  =1  \qquad \forall \alpha\in  \Pur\St_1(\rA) \, .\] 
\end{theorem}
The proof is rather elaborate. The two main steps are
\begin{enumerate}
\item proving that every pure normalized effect $a$ \emph{identifies} a pure state $\alpha$,~meaning that $(a|\rho)=1$ if and only if $\rho  = \alpha$.  
\item proving that, if two pure effects   identify the same state, then  they must coincide. 
\end{enumerate}  
The second step uses Pure Steering in an essential way, suggesting that the distinguishability features of quantum theory are deeply connected with its correlation features.

\subsubsection{The informational dimension}
An easy consequence of the state-effect duality is that every two pure normalized effects are connected by a reversible transformation, just like the pure states. 
   In turn, this leads to a useful result  
   \begin{prop}
   For a given system $\rA\in\Sys$, all pure maximal sets have the same cardinality.  
   \end{prop} 
 The proof idea is simple:   let $\bs a  =  \{a_x\}_{x\in\rX} $ be the measurement that distinguishes among the states in a pure maximal set $\rS=\{\alpha_x ~|~x\in \rX\}$.  As we already observed, all the effects in $\bs a$ must be pure. Since every two pure normalized effects are connected by a reversible transformation, we must have    $ a_{x}   =   a  \,\circ\,  \tU_{x}   \,  \forall x\in\rX$,  where $a$ is fixed (but otherwise arbitrary) effect in  $\Pur\Eff_1  (\rA)$ and $\tU_{x}$ is a reversible transformation.   Applying the effects to the invariant state $\chi$ we then obtain  
 \[(a_x  |\chi)   =  (a|\chi)  \qquad \quad \forall x\in\rX  \, ,\]
and summing over $x$ we get  the equality   $  1  =  |    \rX  |       \,   (a |\chi) $.    Hence, the cardinality of the maximal set $\rS$  is   $| \rS|  \equiv |\rX|  =  1/(a|\chi)$.  Since  $\rS$ is a generic  pure maximal set, we proved the desired result.

In the following, the cardinality of the pure maximal sets in $\rA$  be denoted by $d_\rA$.  We  call it the \emph{informational dimension}, because it is   the number of distinct classical messages that can be encoded in system $\rA$ and decoded without error.    In Quantum Theory, $d_\rA$   is the dimension of the Hilbert space associated to system $\rA$. 

For composite systems, the informational dimension has the product form:
\begin{prop}\label{dprod}
For every pair of systems $\rA$ and $\rB$ one has $d_{\rA\otimes \rB}   =   d_\rA  \,  d_\rB $.  
\end{prop} 
The reason is simply that the product of two pure maximal sets for systems $\rA$ and $\rB$ is a pure maximal set for $\rA\otimes \rB$: it is pure, because the product of two pure states is pure (by Local Tomography) and it is maximal because the product of two internal states is internal (again, by Local Tomography)---hence, maximality follows by proposition \ref{prop:maximal}.  

\subsubsection{The spectral theorem} 

An important  consequence of the state-effect duality  is the ability to decompose every state as a mixture of perfectly distinguishable pure states.  The crucial step is to prove such a decomposition  for the invariant state:
\begin{lemma}\label{crucial}
For every pure maximal set $\{\alpha_x\}_{x=1}^{d_\rA}  \subset  \Pur\St(\rA)$ one has  $\chi       =   \frac 1 {d_\rA}  \sum_{x=1}^{d_\rA} \, \alpha_x$. 
\end{lemma}
This result is extremely important, because it helps us to  cope with the existence of different maximal sets of pure states. To begin with, it allows us to  prove the  analogue of the spectral theorem: 
\begin{theorem}[Spectral Decomposition]\label{th:spec}
  For every vector $v  \in\Str(\rA)$ there exists a pure  maximal set    $\{\alpha_x\}_{x=1}^{d_\rA}  \subset  \Pur\St(\rA)$  and a set of real coefficients  $\{c_x\}_{x=1}^{d_\rA}$ 
    such that
  \begin{equation}
    v=\sum_{x=1}^{d_\rA} \,  c_x\,  \alpha_x \, .
  \end{equation}
Similarly, for every vector $w\in\Effr(\rA)$ there exists a pure discriminating measurement  $\{a_x\}_{x=1}^{d_\rA}$   and a set of real coefficients $\{d_x \}_{x=1}^{d_\rA}$ such that
  \begin{equation}
    w=\sum_{x=1}^{d_\rA}   d_x\,   a_x \, .
  \end{equation}
\end{theorem}

\subsubsection{Orthogonal faces}

Thanks to spectrality, it is easy to retrieve the basic structures of quantum logic.   In general, the faces of a convex set  $C$  form a bounded lattice, with partial order $\preceq$ corresponding to set-theoretic inclusion and with  meet and join operations defined as $F  \wedge G :  =   F  \cap G$ and  $F  \vee G:  =  \bigcap \{  H~|~   F \subseteq H \, , \,  G \subseteq H\}$, respectively.  The lattice is bounded, with the convex set $C$ being the top element and the empty set $\emptyset$ being the bottom element. 
  Hence, the set of deterministic states $C_\rA   :  =  \Det\St (\rA)$ in a convex theory can be seen as a lattice in the above way.   However, our axioms imply much more:  according to them, the faces of the state space form an \emph{orthomodular lattice},~i.~e.~a lattice with an operation of orthogonal complement $\perp$ satisfying the orthomodularity condition $F  \preceq  G  \Longrightarrow    G  =   F   \vee (  G\wedge F^\perp )$. 
  
Let us see why this is the case.  For a given face  $F\subseteq  C_\rA$ we can pick a set  of perfectly distinguishable pure states $\rS_F  = \{ \alpha_x\}_{x=1}^{d_F}  \subset F$ that is \emph{maximal in $F$}, meaning that   no other state in $F$ can be distinguished perfectly from the states in $\rS_F$.    Then, we can define the \emph{barycenter of  $F$} as  
\begin{align}\label{chif}
  \omega_F  :  =  \frac 1 {d_F}  \,     \sum_{x=1}^{d_F} \,  \alpha_x  \, .
  \end{align}    
Since the face $F$ can be compressed into the state space of a smaller system, lemma \ref{crucial} guarantees that the definition of the state $\omega_F$ depends only on $F$, and not  on the maximal set $\rS_F$.    In other words, Eq. (\ref{chif}) sets up a one-to-one correspondence between   faces and their barycenters.  

Now, we can  extend the set $\rS_F$ to a pure maximal set for system $\rA$, say $\{\alpha_x\}_{x=1}^{d_\rA}$.  Let us define the set $\rS_{F^\perp} :  =  \{\alpha_x\}_{x=d_F+1}^{d_\rA}$ and denote by  $F^\perp$  the smallest face containing $\rS_{F^\perp}$.     By construction, it is easy to verify that the set $\rS_{F^\perp}$ is maximal in $F^\perp$ and therefore 
\begin{align*} \omega_{F^\perp}   =  \frac1{d_\rA  -  d_F}   \,  \sum_{x=d_F+ 1}^{d_\rA}   \,  \alpha_x  \, .\end{align*}

$F^\perp$ can be equivalently characterized as the face containing all the  states that are perfectly distinguishable from $F$.  Moreover, it is not hard to show that  
\begin{enumerate}
\item $F  \vee  F^\perp  =  C_\rA$
\item  $F  \wedge F^\perp  = \emptyset$
\item $\left(F^{\perp}\right)^{\perp}  \equiv F$
\item  $F\preceq G   \Longrightarrow    G^\perp  \preceq F^\perp$
\item  $ F  \preceq G  \,  \Longrightarrow \,  G    =    F   \vee  \left( G  \wedge F^{\perp}   \right) $,     
  \end{enumerate} 
where the last two properties are proven by picking a pure maximal set for $F$, extending it to a pure maximal set for $G$, and extending the latter to a pure maximal set for the whole convex set $C_\rA$.   
Properties 1-4  show that the operation $\perp$ is an \emph{orthogonal complement}, while property 5 is the orthomodularity condition.   
Hence, we obtained that the set of faces must be an orthomodular lattice.

\subsubsection{Orthogonal effects}
By the state-effect duality, we can associate every face $F$ with an effect  $a_F$, defined as 
\begin{align}
a_F  :  = \sum_{x=1}^{d_F}    \alpha_x^\dag  \, ,  
\end{align}
where $\rS_F  =\{  \alpha_x\}_{x=1}^{d_F}$ is a pure maximal set in $F$.  Again, it is easy to see that the definition of $a_F$ is independent of the choice of maximal set $\rS_F$.   Indeed, by definition one has $a_F  +  a_{F^\perp}   =   e_\rA$   for every pure maximal set $S_{F^\perp}$.  Varying $\rS_F$ without varying $S_{F^\perp}$ shows  
that the definition of $a_F$ depends only on $F$.  

Thanks to the spectral theorem,  $a_F$ can be operationally characterized the only  effect that happens with unit probability on   $F$ and with zero probability on $F^\perp$:
\begin{prop}\label{prop:identifying}  
 $a_F$ is the unique effect  $a  \in  \Eff(\rA)$ satisfying the conditions 
\begin{align*} 
 (a|  \rho)  &=  1  \qquad \forall \rho  \in  F  \\
 (a|\sigma)  &  =  0   \qquad \forall  \sigma  \in  F^\perp \, .
 \end{align*}
\end{prop}
For this reason, we call $a_F$ the \emph{identifying  effect} of the face $F$.   The set of identifying effects inherits the structure of orthomodular lattice from the set of faces, via the following definitions
\begin{enumerate}
\item $a_F  \preceq  a_G$  iff $F\preceq G$, 
\item $a_F  \wedge  a_G   :  =  a_{F\wedge G}$,
\item $a_F  \vee a_G   :  =  a_{F \vee G}$, and 
\item $a_F^\perp  := a_{F^\perp} $. 
\end{enumerate}    
  In quantum theory, the lattice of identifying effects is the lattice of projectors on subspaces of the Hilbert space.   It is easy to see that the partial order $\preceq$  coincides with the partial  order  $\leq$ induced by the probabilities, namely $a_F   \preceq  a_G $  if and only if $  (a_F|\rho)   \le (a_G|\rho)$ for every state $\rho$.

\subsubsection{Orthogonal projections}
Faces of the state space can also be associated with physical transformations, in the following way:
\begin{definition}
A transformation $\Pi_F \in \Transf(\rA\to \rA)$ is an \emph{orthogonal projection} on the face $F\subseteq C_\rA$  iff the following conditions are satisfied\footnote{In the original work \cite{deri}, we also required that  projections  be \emph{pure}. However, in the context of our axioms, purity is implied by the two conditions in the present definition. A sketch of proof is the following: First, one can prove that for every pure state $\alpha \in  F$ one must have $(\alpha^\dag |  \, \Pi_F  =  (\alpha^\dag |$ (this follows from the definition and from proposition \ref{prop:identifying}).  As a consequence, one also has $(a_F |  \,  \Pi_F  =  (a_F|$. This implies that, for every state $\rho\in\St(\rA)$, the unnormalized state $\Pi_F \,  |  \rho  )$  is proportional to a state in $F$.   Now, for two projections $\Pi_F$ and $\Pi_F'$ one must have   
\begin{align}
(\alpha^\dag|   \,  \Pi_F   |  \rho )     =   (\alpha^\dag  |  \rho)   =   (\alpha^\dag|   \,  \Pi'_F   |  \rho )  \, ,
\end{align}
for every pure state $\alpha\in F$.    Since the states $\Pi_F\,  |\rho)$ and $\Pi_F' \, |\rho)$ are proportional to states in $F$ and $\alpha\in F$ is a generic pure state, Ideal Compression implies $\Pi_F\,  |\rho)  =  \Pi_F' \, |\rho)$, or equivalently, $\Pi_F  =  \Pi_F'$, because the state $\rho$ is generic.      }  

\begin{align}\label{projection1}
\begin{aligned}
\Qcircuit @C=1em @R=.7em @! R {
 &\prepareC{ { \rho}}&\poloFantasmaCn{\rA} \qw &  \gate{{\Pi_F}}  &\poloFantasmaCn{\rA} \qw   & \qw     }
\end{aligned}   \quad &=   \begin{aligned}
\Qcircuit @C=1em @R=.7em @! R {
 &\prepareC{ { \rho}}&\poloFantasmaCn{\rA} \qw &\qw  }
\end{aligned}     \qquad \quad \forall \rho  \in  F  \\   
\nonumber  & \\
\label{projection2}
\begin{aligned}
\Qcircuit @C=1em @R=.7em @! R {
 &\prepareC{ { \sigma}}&\poloFantasmaCn{\rA} \qw &  \gate{{\Pi_F}}  &\poloFantasmaCn{\rA} \qw   & \qw     }
\end{aligned}    \quad  &=  \quad 0   \qquad \qquad\qquad \, \forall \sigma  \in  F^\perp \, .  
\end{align}   
\end{definition}
The definition is non-empty:  thanks to Purification and Purity of Composition, we are able to construct a \emph{pure} projection $\Pi_F$ for every face  $F$. 
Moreover, it follows from the definition that the  projection $\Pi_F$ is  unique.  
% The fact that projections are pure is essential: thanks to Pure Composition, we have that every projection sends a pure state into (a multiple of) a pure state.  This fact will be essential for the derivation of the Hilbert space framework, where we will use projections to extend the characterization of two-dimensional systems to higher-dimensional ones.    

In addition to purity, projections have a number of properties, including
\begin{enumerate}
\item  $ (a_F^\perp|   \,  \Pi_F   =    0  $
\item  $  (a_G|   \,  \Pi_F  =  (a_G|$  whenever $G\preceq F$     
\item   for every input state $\rho$,  the normalized output state  $\tau : =   \Pi_F \, |\rho)/ (e_\rA | \, \Pi_F  \,|\rho)$ belongs to $F$
\item  $  \Pi_G \,  \Pi_F   =  \Pi_F \Pi_G  = \Pi_G$  whenever $G\preceq F$.
\end{enumerate}

\subsection{Interaction between correlation and distinguishability structures}  
We have seen that our axioms imply peculiar features, both in the way systems correlate and in the way states can be distinguished.   It is time to combine these two types of features and to explore the consequences. 
  
\subsubsection{The Schmidt bases}
  
Combining Pure Steering and Spectral Decomposition, we are now in position to give the operational version  of the Schmidt bases in quantum theory.  The result can be summarized as follows: 
\begin{prop}
Let $\Psi$ be a pure state of $\rA\otimes \rB$ and let $\rho_\rA$ and $\rho_\rB$ be its marginals on systems $\rA$ and $\rB$, respectively.  Then, for every spectral decomposition  
\begin{align*}
\rho_\rA    =  \sum_{x=1}^r  \,   p_x  \,  \alpha_x    \, ,
\end{align*}
there exists a set of perfectly distinguishable pure states $ \{\beta_x \}_{x= 1}^r \subset \Pur\St(\rB)$ such that  
\begin{align}
\rho_\rB  =  \sum_{x= 1}^r  \,  p_x     \,  \beta_x  \, .  
\end{align}
Moreover, 
one has
\begin{align}
    \begin{aligned}
    \Qcircuit @C=1em @R=.7em @! R {\multiprepareC{1}{\Psi}&\poloFantasmaCn{\rA}\qw&\measureD{a_x}\\
      \pureghost{\Psi}&\poloFantasmaCn{\rB}\qw&\measureD{b_{y}}}
  \end{aligned}    =  \quad  
 \left\{   \begin{array}{ll}  
  p_x  \,   \delta_{xy}  \qquad &  x,y\in  \{1,\dots, r\}  \\
  &\\
  0   \qquad &  x,y  \not \in  \{1,\dots, r\}   
  \end{array}
  \right.  
  \end{align}
  for every two measurements $\bs a  =  \{a_x\}_{x  =1}^{k_\rA}$ and $\bs  b  =  \{b_y\}_{y=1}^{k_\rB}$  satisfying $a_x =  \alpha_x^\dag$ and $b_y   =  \beta_y^\dag$ for every $x\le r$ and for every $y\le r$. 
  \end{prop}
  In particular, applying the result to the Bell state $\Phi$, we obtain   that the invariant state $\chi_{\overline A}$ can be decomposed as  $\chi_{\overline A}   =  \frac 1 {d_\rA}    \,  \sum_{x=1}^{d_\rA} \,  \overline \alpha_x$, for a suitable set of perfectly distinguishable pure states $\{\overline \alpha_x\}_{x=1}^{d_\rA}$.  
  In particular, this implies that conjugate systems have the same informational dimension:  
  \begin{cor}
  For every system $\rA$, one has $d_{\over \rA} =  d_\rA$.  
  \end{cor}
Combined with the fact that the informational dimension is multiplicative (proposition \ref{dprod}), the above result implies that the composite system $\rA\otimes \over \rA$ has informational dimension    
\[d_{\rA\otimes  \over \rA}   =  d_\rA^2 \, .\] 
\subsubsection{The maximum probability of conclusive teleportation}  

In our construction of conclusive teleportation, the teleportation probability  was equal to the probability of the state $ \Phi$ in an ensemble decomposition of the invariant state $\chi_\rA\otimes \chi_{\overline \rA}$, cf. Eq. (\ref{probtele}).   Now, since $\chi_\rA\otimes \chi_{\overline \rA}$ is the invariant state, it can be decomposed as
\[    \chi_\rA\otimes \chi_{\overline \rA}    = \frac1{d_\rA^2} \,    \sum_{x=1}^{d_\rA^2}   \,  \Phi_x \, ,   \]    
for every  pure maximal set $\{  \Phi_x\}_{x=1}^{d_\rA^2}$.  
The maximum probability of the Bell state in a convex decomposition of $\chi_\rA\otimes \chi_{\over \rA}$ is then given by 
  \begin{align}\label{pmax}  p^{\max}_\rA  =  \frac 1 {d_A^{2}}  \, .  \end{align} 
Inserting the above equality into the teleportation upper bound (\ref{teleupper})  we obtain the relation 
  \begin{align}\label{Dada}
  D_\rA  \le d_\rA^2  \, .
 \end{align}

 In the next paragraph we will see how to obtain the converse inequality.

%In quantum theory the bound is saturated:  $d_\rA$ is the dimension of the Hilbert space of system $\rA$ and $D_\rA$ is the dimension of the space of  $d_\rA\times d_\rA$ Hermitian matrices, which is indeed equal to $d_\rA^2$.  To prove the saturation of the bound from the principles, we need to 
\subsubsection{The teleportation lower bound}

Thanks to the state-effect duality, it is possible to establish a lower bound on the state space dimension. The proof is a little bit laborious and consists of two steps: 
\begin{enumerate}
\item show that the effect   $\Phi^\dag$ that identifies the Bell state is  of the form 
\begin{align*}
\begin{aligned}
\Qcircuit @C=1em @R=.7em @! R {
   &  \qw \poloFantasmaCn{\rA}   &  \multimeasureD{1}{\Phi^\dag}  \\
 &   \qw \poloFantasmaCn{\over \rA}  &  \ghost{\Phi^\dag}}
\end{aligned}     =  \quad 
\begin{aligned}
\Qcircuit @C=1em @R=.7em @! R {
   &  \qw \poloFantasmaCn{\rA}   &     \gate{\tU}  &     \qw \poloFantasmaCn{\rA}   &  \multigate{1}{\tS_{\rA,\over \rA}}  &    \qw \poloFantasmaCn{\over \rA}   &  \multimeasureD{1}{E}  \\
 &  \qw \poloFantasmaCn{\over \rA}   &     \qw  &     \qw    &  \ghost{\tS_{\rA,\over\rA}}  &    \qw \poloFantasmaCn{\rA}   &  \ghost{E}}
 \end{aligned}
\end{align*}   
where $E$ is the  effect achieving maximum teleportation probability,  $\tS_{\rA,\over\rA}$ is the swap, and  $\tU$ is some reversible transformation.  
\item show that, with a suitable choice of basis   for the vector space $\St_\R(\rA)$, every reversible transformation   $\tU$ is represented by an orthogonal matrix    $M_\tU$.  
\end{enumerate}  
Once these two results are established, we can expand  the Bell state $\Phi$ and the teleportation effect  $E$ as in Eq. (\ref{PhiE}), thus obtaining 
\begin{align} 
  1  =  (\Phi^\dag |  \Phi  )   =  \Tr  [ \Phi  \, E   \,  M_\tU   ]  =   p^{\max}_\rA  \,    \Tr[M_\tU]    \le    p^{\max}_\rA  \,    D_\rA  \, ,
 \end{align} 
 having used the teleportation equality $  \Phi \, E   =  p^{\max}_\rA  \,  I_{D_\rA}$ and the fact that the trace of an orthogonal matrix cannot be larger than the trace of the identity.  Hence, we obtained the \emph{teleportation lower bound}  
 \begin{align}
 D_\rA  \ge  \frac {1}{p^{\max}_\rA}   \, .
  \end{align}
  
Combining the teleportation lower bound   with Eqs. (\ref{pmax}) and (\ref{Dada}), we  obtain the equality   
 \begin{align}\label{Dada2}
D_\rA   =   d_\rA^2 \, .
\end{align}  

\subsection{Qubit structures}  
So far, we avoided giving a concrete representation of our state spaces: all the quantum features that we have shown followed \emph{directly} from the principles.   
We now proceed to analyze  some features that are more closely related to the concrete geometrical shape of the quantum state spaces. We will first see that all two-dimensional systems  in our theory have qubit state spaces.  Leveraging on this fact, we will then derive two features of higher-dimensional systems:  \emph{i)} an operational version of the superposition principle, and \emph{ii)} the fact that all systems of the same dimension are operationally equivalent.  
  
\subsubsection{Derivation of the qubit}  

Showing that the states of a two-dimensional system can be described by density matrices is quite easy.  This can be done geometrically, by  showing that the deterministic states form a 3-dimensional Euclidean ball.  The 3-dimensionality is obvious from the equality $D_\rA  =  d_\rA^2$, which for $d_\rA  = 2$  implies that the convex set $C_\rA  =  \Det\St(\rA)$ is a three-dimensional manifold  \footnote{In general,  the dimension of the convex set $C_\rA$ is given by $D_\rA-1$.}.  
Then, we can make a simple geometrical reasoning: 

\begin{enumerate}
\item  all the pure states are generated from a fixed pure state by application of reversible transformations, and, by choosing a suitable basis for the state space, such transformations act in the 3-dimensional space as orthogonal matrices.  
\item  all states on the border of $C_\rA$ are pure---otherwise, Perfect State Discrimination and proposition  \ref{prop:onlypure} would imply  $d_\rA >2$.  This means that, if we move away from the invariant state $\chi_\rA$ in an arbitrary direction, at some point we will hit a pure state.  
\end{enumerate}
 In the ordinary 3-dimensional space, the sphere  is the  only (closed) 3-dimensional convex set  generated by orthogonal matrices and with only pure states on the border.   

Once we established that the convex set $C_\rA$ is a sphere, we can represent every normalized state $\rho\in C_\rA$ with a  density matrix  $S_\rho$.  In particular, the pure states will be of the form  
\begin{align}\label{qubitpure}
S_\alpha  = \begin{pmatrix}   p    &   \sqrt{p(1-p) } \, e^{-i\theta}  \\
\sqrt{p(1-p) } \, e^{i\theta}  &     1-p   
\end{pmatrix}    =       |\alpha\>\<\alpha| \\  \nonumber \\
\nonumber   |\alpha\>  :  =   \sqrt{p} \,  |0\> +  e^{i\theta}\, \sqrt{1-p} \, |1\>     \, ,
\end{align}
for some probability $p\in [0,1]$ and some phase $\theta\in  [0,2\pi)$.   Once we have chosen this representation, it is obvious that every effect $a \in  \Eff (\rA)$ must be described by a positive semidefinite matrix $E_a$   upper bounded by the identity and that probabilities are given  by the Born rule  
\begin{align}
(a|\rho)   =  \Tr  [   E_a  \,  S_\rho] \, . 
\end{align}
Moreover, the state-effect duality imposes  that \emph{all} such matrices represent valid effects.    

\iffalse
Finally, the fact that all pure states are connected by reversible transformations  implies that the group of reversible transformations is either $\mathbb{SO}  (3)$ or the whole $\mathbb{O} (3)$.  A simple argument allows to exclude $\mathbb {O}  (3)$, because the inversion would correspond to the universal NOT, which is not an admissible transformation because it does not respect the positivity of probabilities when applied on entangled states.   Hence, we remain with $\mathbb{SO}(3)$,  meaning that the action of a reversible transformation $\tU$ is given by 
\begin{align}
S_{\tU  \rho}  =  U  \,  S_\rho  \,  U^\dag   \qquad \quad U  \in    \mathbb{SU}  (2) \, . 
\end{align} 
\fi

\subsubsection{The superposition principle}

Pure states in quantum theory satisfy the so-called ``superposition principle", which just means that they are in one-to-one correspondence with the rays of the underlying Hilbert space.   
\emph{Per se}, this statement has hardly any operational meaning.   However, one can formulate an operational version in general OPTs: 
\begin{definition}[Superposition Principle]
We say that system $\rA$ \emph{satisfies the superposition principle} iff for every pure maximal set $\rS =\{  \alpha_x~|~  x  \in  \rX \}  \subset \Pur\St(\rA)$ and for every probability distribution $\{p_x\}_{x\in\rX}$ there exists one pure state $  \psi $ such that  
\begin{align}
 \begin{aligned}
    \Qcircuit @C=1em @R=.7em @! R {\prepareC{\psi}&\poloFantasmaCn{\rA}\qw&\measureD{a_x}}  
    \end{aligned}
     =    \quad p_x  \qquad  \forall x\in\rX  \, , 
\end{align}
for every measurement $\bs  a  = \{  a_x\}_{x\in\rX}$ that perfectly distinguishes among the states in the maximal set $\rS$.  
\end{definition}

Now, in a theory satisfying our principles we know that the two-dimensional systems are quantum---and therefore satisfy the superposition principle.  Thanks to Ideal Compression, it is then easy to generalize the result to systems of arbitrary dimension: given two perfectly distinguishable pure states, one can encode them into a two-dimensional system, use the Bloch sphere representation to find the superposition state, and come back with the decoding operation.  Iterating this procedure, we can superpose any number of perfectly distinguishable pure states.   

As a simple application of the superposition principle, we obtain  the following  
\begin{prop}
A  state $\rho_\rA$  with spectral decomposition $\rho_\rA  =  \sum_{x=1}^r  \,   p_x \,  \alpha_x$  has a purification with purifying system $\rB$ if and only if  $d_\rB  \ge r$. 
\end{prop} 
The ``only if" part was already clear from the Schmidt decomposition.   For the ``if" part, it is enough to pick $r$ perfectly distinguishable pure states of $\rB$, say $\{\beta_x\}_{x=1}^r$, and to  superpose the product states $\{\alpha_x \otimes \beta_x\}_{x=1}^r$ with probabilities $\{p_x\}_{x=1}^r$.    The resulting pure state $\Psi  \in  \Pur\St(\rA\otimes \rB)$ is the desired purification.  

\subsubsection{The superposition principle for transformations}
The superposition principle allows us to glue distinguishable states in any way we like.     Thanks to the state-transformation isomorphism, we can extend this idea to transformations.  For example, consider a set of pure transformations $\{   \tA_x  ~|~x\in\rX \}  \subset  \Pur\Transf(\rA\to \rB)$  and suppose that they have \emph{orthogonal support}, that is, that there exists a set of orthogonal faces $\{  F_x~|~x\in\rX\}$ such that  
\begin{align}
\tA_x    =    \tA_x   \,  \Pi_{F_x}  \qquad \forall x\in\rX  \, .
\end{align}
Then, it is possible to find a pure transformation  $\tA  \in \Pur\Transf(\rA\to \rB)$ such that 
\begin{align}
\tA  \, \Pi_{F_x}     =  \tA_x  \quad \forall x\in \rX  \, .   
\end{align}
The result follows by noticing that the Choi states $\{  \Phi_{\tA_x}~|~x\in\rX\}$ are  proportional to pure and perfectly distinguishable states and by applying the superposition principle to corresponding the normalized states.

\subsubsection{Equivalence of pure maximal sets up to reversible transformations}
Using the superposition principle for transformations  we can prove that all pure maximal sets of the same cardinality are equivalent:
\begin{prop}
Let $\{\alpha_x\}_{x=1}^{d_\rA}$ and $\{\beta_y\}_{y=1}^{d_\rB} $be  pure maximal sets for systems $\rA$ and $\rB$, respectively.  If $d_\rA  =  d_\rB$, then there exists a reversible transformation $\tU  \in \Transf(\rA\to \rB)$ such that 
\begin{align*}
\begin{aligned}
    \Qcircuit @C=1em @R=.7em @! R {\prepareC{\alpha_x}&\poloFantasmaCn{\rA}\qw&\gate{\tU} &\poloFantasmaCn{\rB}\qw  &\qw    }  
    \end{aligned}
 =  \quad
\begin{aligned}
    \Qcircuit @C=1em @R=.7em @! R {\prepareC{\beta_x}&\poloFantasmaCn{\rB}\qw& \qw    }  
    \end{aligned}   \qquad \forall x\in\rX \, .
\end{align*}
\end{prop}   
The result follows immediately from the application of the superposition principle to the pure transformations $\tA_x  =    |\beta_x  )(  \alpha_x^\dag| $.  
As a corollary, we have that all systems of the same dimension are operationally equivalent.

\subsection{The density matrix}

We finally reached to the end of the reconstruction. It is now time to enter into the specific details of the Hilbert space formalism of quantum theory.    Our strategy to reconstruct the Hilbert space formalism is to show that, for every system $\rA$, there exists a one-to-one linear map from the vector space $\St_\R (\rA)$ to the space of $d_\rA\times d_\rA$ Hermitian matrices, with the property that  the convex set of deterministic states  is mapped to the convex set of density matrices (non-negative matrices with unit trace).   

Let us see how this can be proven. Since the dimension of the state space satisfies the relation $D_\rA  =  d_\rA^2$,   every  vector $  v  \in  \Str(\rA)$ can be represented as  square $d_\rA\times d_\rA$ real matrix $
M_v$.    In turn, the matrix $M_v$  can be turned into a complex Hermitian matrix  $S_v$, applying the linear transformation 
\begin{equation}
  S_v:=\left( M_v+  M_v^T\right)+i  \left( M_v- M_v^T\right),
\end{equation}
where $M^T$ denotes the transpose of $M$.  The problem is now to find a suitable representation in which normalized states $\rho  \in C_\rA$ correspond to density matrices, that is $S_\rho  \ge 0 $ and $\Tr [\rho]  =1$.   
To find such a representation, we follow  Hardy's method \cite{hardy01}:  we pick a pure maximal set $\{  \alpha_m\}_{m=1}^{d_\rA}$  and define the diagonal elements of the matrix $S_\rho$ as 
\[ [S_{\rho}]_{mm}   :  =  (  \alpha_m^\dag|  \rho)  \, ,\]
In this way, we guarantee the unit-trace condition $\Tr  [S_\rho]  =1$.   To define the off-diagonal elements, we consider the two-dimensional faces  $F_{mn}   :  =  \{\alpha_m\}   \vee  \{\alpha_n\}$,  $n>m$.  Projecting the state inside these faces, we obtain the states  
\[  \left|  \rho^{mn} \right)    =    \frac{\Pi_{F_{mn}}  |\rho)}{ (e_\rA|   \,  \Pi_{F_{mn}}  \,  |\rho)    }  \qquad n>m  \, . \]  
Since every state $\rho^{mn}$ is belongs to a two-dimensional face, it can be encoded into a qubit system and can be associated with a density matrix  $\tau^{mn}$.  The off-diagonal elements $ [S_{\rho}]_{mn}$ and $[S_\rho]_{nm}$ are defined in term of the qubit density matrix $\tau^{mn}$, as 
\[    [S_{\rho}]_{mn}   :  =    [\tau^{mn}]_{01}   \qquad {\rm and} \qquad     [S_{\rho}]_{nm}   :  =  [\tau^{mn}]_{10}  \, .  \]  
The matrix $S_\rho$ defined in this way is clearly Hermitian and, with a little bit of work, one can see that the linear map $\rho  \mapsto  S_\rho$  is one-to-one.

At this point the problem is to guarantee that the matrix $S_\rho$ is positive.    We consider first the case of pure states  $\alpha\in \Pur\St(\rA)$, for which one has  
\[     [  S_\alpha  ]_{mn}   =    \sqrt{p_m \, p_n}  \,   e^{i  \theta_{mn}}   \]
  where $\{  p_m\}_{m=1}^{d_\rA}$ is a suitable probability distribution and $\{  \theta_{mn}   \}$ are phases satisfying the conditions $\theta_{mm}  =0$ for every $m$ and $\theta_{nm}  =  - \theta_{mn}$ for every $n>m$. 
This expression follows from the fact that each state $\left | \alpha^{mn} \right) =  \Pi_{F_{mn}}   \,  |\alpha)  / (e_\rA| \Pi_{F_{mn}}   \,  |\alpha) $ is pure and, once encoded into a qubit, it has a density matrix of the form  (\ref{qubitpure}).   
In order to prove positivity, we need to show that the phases $\theta_{mn}$ are of the form  $\theta_{mn}  =  \gamma_m  -  \gamma_n$, for some phases $\{\gamma_m\}$.  The strategy is to prove the result first in dimension $d_\rA  = 3$ and then to extend it to arbitrary dimensions.  

Once we have proven that pure states correspond to rank-one projectors, it remains to show that \emph{all} such projectors correspond to pure states.  This can be done by using the superposition principle (both for states and for reversible transformations).   
Having proven that the set of pure states is in one-to-one correspondence with the set of rank-one projectors, it follows by convexity that the set of states is in one-to-one correspondence with the set of density matrices.   In short, all state spaces are quantum.

To complete our reconstruction, we invoke  theorem \ref{th:staspec}, which guarantees that  the tests  in our theory are in one-to-one correspondence with the test allows by quantum theory.   

\section{Conclusions}\label{sec:conclusions}

Quantum theory can be rebuilt from bottom to top starting from six basic principles.   The principles do not refer to specific physical systems such as particles or waves: instead, they are the rules that dictate how information can be processed. 
%We formulated these rules in a framework of general theories of information, regarded as extensions of probability theory, and, in turn, as  extensions of logic.   
The first five principles---Causality, Purity of Composition, Local Tomography, Perfect State Discrimination, and Ideal Compression---can be thought of as  requirements for a standard theory of information. 
On the background of these five principles,  the sixth---Purification---stands out as {\em the} quantum principle, which brings in counterintuitive features like entanglement, no cloning, and teleportation. Purification gives the agent the power to harness randomness, by simulating the preparation of every state through the preparation of a pure bipartite state. When this is done, the agent has an intrinsic guarantee that no side information can hide outside her control.    
The moral of our reconstruction is quantum theory is the standard theory of information that allows for maximal control of randomness.

%From a fundamental point of view, it is now interesting to address the
%same issues reaised by Quantum Theory in the general context of of
%probabilistic theories, e.~g.~non-locality, contextuality, and the
%existence of a local realistic interpretation, as in Ref.\cite{francux}.

%From an information-theoretic point of view, one can explore the
%consequences of relaxing any of the postulates. Most interestingly,
%non-causal theories can be the playground for new kinds of computation
%models, as explored in Refs.~\cite{vali,brukner,giulio,timo}.

\subparagraph*{Acknowledgements}     

The work is supported by the Templeton Foundation under the project ID\# 43796 \emph{A Quantum-Digital Universe},  by the Foundational Questions Institute through the large grant   \emph{The fundamental principles of information dynamics}  (FQXi-RFP3-1325), by the 1000 Youth Fellowship Program of China,  and by the National Natural Science Foundation of China through Grants 11450110096 and 11350110207.
GC acknowledges  the hospitality of the Simons Center for the Theory of Computation and of Perimeter Institute for Theoretical Physics.  Research at Perimeter Institute for Theoretical Physics is supported in part by the Government of Canada through NSERC and by the Province of Ontario through MRI.

\bibliography{bibliography}

\begin{thebibliography}{10}
\expandafter\ifx\csname url\endcsname\relax
  \def\url#1{\texttt{#1}}\fi
\expandafter\ifx\csname urlprefix\endcsname\relax\def\urlprefix{URL }\fi
\expandafter\ifx\csname href\endcsname\relax
  \def\href#1#2{#2} \def\path#1{#1}\fi

\bibitem{puri}
G.~Chiribella, G.~M. D'Ariano, P.~Perinotti,
  \href{http://link.aps.org/doi/10.1103/PhysRevA.81.062348}{Probabilistic
  theories with purification}, Phys. Rev. A 81 (2010) 062348.
\newblock \href {http://dx.doi.org/10.1103/PhysRevA.81.062348}
  {\path{doi:10.1103/PhysRevA.81.062348}}.
\newline\urlprefix\url{http://link.aps.org/doi/10.1103/PhysRevA.81.062348}

\bibitem{deri}
G.~Chiribella, G.~M. D'Ariano, P.~Perinotti,
  \href{http://link.aps.org/doi/10.1103/PhysRevA.84.012311}{Informational
  derivation of quantum theory}, Phys. Rev. A 84 (2011) 012311.
\newblock \href {http://dx.doi.org/10.1103/PhysRevA.84.012311}
  {\path{doi:10.1103/PhysRevA.84.012311}}.
\newline\urlprefix\url{http://link.aps.org/doi/10.1103/PhysRevA.84.012311}

\bibitem{epr}
A.~Einstein, B.~Podolsky, N.~Rosen,
  \href{http://link.aps.org/doi/10.1103/PhysRev.47.777}{Can quantum-mechanical
  description of physical reality be considered complete?}, Phys. Rev. 47
  (1935) 777--780.
\newblock \href {http://dx.doi.org/10.1103/PhysRev.47.777}
  {\path{doi:10.1103/PhysRev.47.777}}.
\newline\urlprefix\url{http://link.aps.org/doi/10.1103/PhysRev.47.777}

\bibitem{schrodinger1935discussion}
E.~Schr{\"o}dinger, Discussion of probability relations between separated
  systems, in: Mathematical Proceedings of the Cambridge Philosophical Society,
  Vol.~31, Cambridge Univ Press, 1935, pp. 555--563.

\bibitem{birkvon}
G.~Birkhoff, J.~V. Neumann, \href{http://www.jstor.org/stable/1968621}{The
  logic of quantum mechanics}, Annals of Mathematics 37~(4) (1936) pp.
  823--843.
\newline\urlprefix\url{http://www.jstor.org/stable/1968621}

\bibitem{mackey}
G.~W. Mackey, Quantum mechanics and hilbert space, American Mathematical
  Monthly (1957) 45--57.

\bibitem{ludwig}
G.~Ludwig, Versuch einer axiomatischen grundlegung der quantenmechanik und
  allgemeinerer physikalischer theorien, Zeitschrift f{\"u}r Physik 181~(3)
  (1964) 233--260.

\bibitem{piron}
C.~Piron, Axiomatique quantique, Helvetica physica acta 37~(4-5) (1964) 439.

\bibitem{jauchpiron}
J.~Jauch, C.~Piron, On the structure of quantal proposition systems, in: The
  Logico-Algebraic Approach to Quantum Mechanics, Springer, 1975, pp. 427--436.

\bibitem{beltrametticassinelli}
E.~G. Beltrametti, G.~Cassinelli, The logic of quantum mechanics, Vol.~15,
  Cambridge University Press, 2010.

\bibitem{bobquantumlogic}
B.~Coecke, D.~Moore, A.~Wilce, Current research in operational quantum logic:
  algebras, categories, languages, Vol. 111, Springer Science \& Business
  Media, 2000.

\bibitem{wikipedia}
Quantum logic, \url{http://en.wikipedia.org/wiki/Quantum\_logic}, accessed:
  2015-04-30.

\bibitem{bb84}
C.~H. Bennett, G.~Brassard, Quantum cryptography: public key distribution and
  coin tossing, in: Proceedings of IEEE International Conference on Computers,
  Systems and Signal Processing, 1984, pp. 175--179.

\bibitem{e91}
A.~K. Ekert, Quantum cryptography based on bell's theorem, Physical review
  letters 67~(6) (1991) 661.

\bibitem{shor}
P.~W. Shor, Algorithms for quantum computation: Discrete logarithms and
  factoring, in: Foundations of Computer Science, 1994 Proceedings., 35th
  Annual Symposium on, IEEE, 1994, pp. 124--134.

\bibitem{grover}
L.~K. Grover, \href{http://doi.acm.org/10.1145/237814.237866}{A fast quantum
  mechanical algorithm for database search}, in: Proceedings of the
  Twenty-eighth Annual ACM Symposium on Theory of Computing, STOC '96, ACM, New
  York, NY, USA, 1996, pp. 212--219.
\newblock \href {http://dx.doi.org/10.1145/237814.237866}
  {\path{doi:10.1145/237814.237866}}.
\newline\urlprefix\url{http://doi.acm.org/10.1145/237814.237866}

\bibitem{wootzu}
W.~Wootters, W.~Zurek, A single quantum cannot be cloned, Nature 299~(5886)
  (1982) 802--803.

\bibitem{dieks1982communication}
D.~Dieks, Communication by epr devices, Physics Letters A 92~(6) (1982)
  271--272.

\bibitem{telep}
C.~H. Bennett, G.~Brassard, C.~Cr{\'e}peau, R.~Jozsa, A.~Peres, W.~K. Wootters,
  Teleporting an unknown quantum state via dual classical and
  einstein-podolsky-rosen channels, Physical review letters 70~(13) (1993)
  1895.

\bibitem{densec}
C.~H. Bennett, S.~J. Wiesner, Communication via one-and two-particle operators
  on einstein-podolsky-rosen states, Physical review letters 69~(20) (1992)
  2881.

\bibitem{littlemore}
C.~A. Fuchs, Quantum mechanics as quantum information, mostly, Journal of
  Modern Optics 50~(6-7) (2003) 987--1023.

\bibitem{cryptofound}
G.~Brassard, Is information the key?, Nature Physics 1~(1) (2005) 2--4.

\bibitem{fuchs2001quantum}
C.~A. Fuchs, et~al., Quantum foundations in the light of quantum information,
  NATO SCIENCE SERIES SUB SERIES III COMPUTER AND SYSTEMS SCIENCES 182 (2001)
  38--82.

\bibitem{fuchs2011coming}
C.~A. Fuchs, Coming of age with quantum information, Coming of Age With Quantum
  Information, by Christopher A. Fuchs, Cambridge, UK: Cambridge University
  Press, 2011.

\bibitem{hardy01}
L.~Hardy, Quantum theory from five reasonable axioms, arXiv preprint
  quant-ph/0101012.

\bibitem{maurofirst}
G.~M. D'Ariano, How to derive the hilbert-space formulation of quantum
  mechanics from purely operational axioms, AIP Conference Proceedings 844
  (2006) 101.

\bibitem{maurobook}
G.~M. D'Ariano, Probabilistic theories: what is special about quantum
  mechanics, in: A.~Bokulich, G.~Jaeger (Eds.), Philosophy of quantum
  information and entanglement, Cambridge University Press, Cambridge, 2010,
  pp. 85--126.
\newblock \href {http://dx.doi.org/10.1017/CBO9780511676550.007}
  {\path{doi:10.1017/CBO9780511676550.007}}.

\bibitem{hardy11}
L.~Hardy, Reformulating and reconstructing quantum theory, arXiv:1104.2066.

\bibitem{masanes11}
L.~Masanes, M.~P. M{\"u}ller, A derivation of quantum theory from physical
  requirements, New Journal of Physics 13~(6) (2011) 063001.

\bibitem{dakic11}
B.~{Dakic}, C.~{Brukner}, {Quantum Theory and Beyond: Is Entanglement
  Special?}, in: H.~Halvorson (Ed.), Deep Beauty: Understanding the Quantum
  World through Mathematical Innovation, Cambridge University Press, 2011, pp.
  365--392.

\bibitem{goyal10}
P.~Goyal, K.~H. Knuth, J.~Skilling,
  \href{http://link.aps.org/doi/10.1103/PhysRevA.81.022109}{Origin of complex
  quantum amplitudes and feynman's rules}, Phys. Rev. A 81 (2010) 022109.
\newblock \href {http://dx.doi.org/10.1103/PhysRevA.81.022109}
  {\path{doi:10.1103/PhysRevA.81.022109}}.
\newline\urlprefix\url{http://link.aps.org/doi/10.1103/PhysRevA.81.022109}

\bibitem{masanes12}
L.~Masanes, M.~P. M{\"u}ller, R.~Augusiak, D.~P{\'e}rez-Garc{\'\i}a, Existence
  of an information unit as a postulate of quantum theory, Proceedings of the
  National Academy of Sciences 110~(41) (2013) 16373--16377.

\bibitem{wilce2012conjugates}
A.~Wilce, Conjugates, correlation and quantum mechanics, arXiv:1206.2897.

\bibitem{barnum2014higher}
H.~Barnum, M.~P. Mueller, C.~Ududec, Higher-order interference and
  single-system postulates characterizing quantum theory, arXiv:1403.4147.

\bibitem{barrett07}
J.~Barrett,
  \href{http://link.aps.org/doi/10.1103/PhysRevA.75.032304}{Information
  processing in generalized probabilistic theories}, Phys. Rev. A 75 (2007)
  032304.
\newblock \href {http://dx.doi.org/10.1103/PhysRevA.75.032304}
  {\path{doi:10.1103/PhysRevA.75.032304}}.
\newline\urlprefix\url{http://link.aps.org/doi/10.1103/PhysRevA.75.032304}

\bibitem{nobroad}
H.~Barnum, J.~Barrett, M.~Leifer, A.~Wilce, Generalized no-broadcasting
  theorem, Physical Review Letters 99~(24) (2007) 240501.
\newblock \href {http://dx.doi.org/10.1103/PhysRevLett.99.240501}
  {\path{doi:10.1103/PhysRevLett.99.240501}}.

\bibitem{barnum08}
H.~{Barnum}, J.~{Barrett}, M.~{Leifer}, A.~{Wilce}, {Teleportation in General
  Probabilistic Theories}, ArXiv e-prints\href {http://arxiv.org/abs/0805.3553}
  {\path{arXiv:0805.3553}}.

\bibitem{hardy2013}
L.~Hardy, A formalism-local framework for general probabilistic theories,
  including quantum theory, Mathematical Structures in Computer Science 23~(02)
  (2013) 399--440.
\newblock \href {http://dx.doi.org/10.1017/S0960129512000163}
  {\path{doi:10.1017/S0960129512000163}}.

\bibitem{barnum11}
H.~Barnum, A.~Wilce, Information processing in convex operational theories,
  Electronic Notes in Theoretical Computer Science 270~(1) (2011) 3--15.

\bibitem{abramsky2004}
S.~Abramsky, B.~Coecke, A categorical semantics of quantum protocols, in: Logic
  in Computer Science, 2004. Proceedings of the 19th Annual IEEE Symposium on,
  IEEE, 2004, pp. 415--425.
\newblock \href {http://dx.doi.org/10.1109/LICS.2004.1319636}
  {\path{doi:10.1109/LICS.2004.1319636}}.

\bibitem{abramsky2008}
S.~Abramsky, B.~Coecke, Categorical quantum mechanics, in: K.~Engesser, D.~M.
  Gabbay, D.~Lehmann (Eds.), Handbook of quantum logic and quantum structures:
  quantum logic, Elsevier, 2008, pp. 261--324.
\newblock \href {http://dx.doi.org/10.1016/B978-0-444-52869-8.50014\-1}
  {\path{doi:10.1016/B978-0-444-52869-8.50014\-1}}.

\bibitem{coecke2010universe}
B.~Coecke, A universe of processes and some of its guises, in: H.~Halvorson
  (Ed.), Deep Beauty: Understanding the Quantum World Through Mathematical
  Innovation, Cambridge University Press, 2010, pp. 129--186.

\bibitem{coecke2011categories}
B.~Coecke, {\'E}.~O. Paquette, Categories for the practising physicist, in: New
  Structures for Physics, Springer, 2011, pp. 173--286.

\bibitem{mac1978categories}
S.~Mac~Lane, Categories for the working mathematician, Vol.~5, Springer Science
  \& Business Media, 1978.

\bibitem{coecke2010quantum}
B.~Coecke, Quantum picturalism, Contemporary physics 51~(1) (2010) 59--83.
\newblock \href {http://dx.doi.org/10.1080/00107510903257624}
  {\path{doi:10.1080/00107510903257624}}.

\bibitem{selinger2011survey}
P.~Selinger, A survey of graphical languages for monoidal categories, in:
  B.~Coecke (Ed.), New Structures for Physics, Vol. 813 of Lecture Notes in
  Physics, 2011.
\newblock \href {http://dx.doi.org/10.1007/978-3-642-12821-9\_4}
  {\path{doi:10.1007/978-3-642-12821-9\_4}}.

\bibitem{chiribella14dilation}
G.~Chiribella, Dilation of states and processes in operational-probabilistic
  theories, in: B.~Coecke, I.~Hasuo, P.~Panangaden (Eds.), {\rm Proceedings
  11th workshop on} Quantum Physics and Logic, {\rm Kyoto, Japan, 4-6th June
  2014}, Vol. 172 of Electronic Proceedings in Theoretical Computer Science,
  Open Publishing Association, 2014, pp. 1--14.
\newblock \href {http://dx.doi.org/10.4204/EPTCS.172.1}
  {\path{doi:10.4204/EPTCS.172.1}}.

\bibitem{SpekkensToy}
R.~W. Spekkens,
  \href{http://link.aps.org/doi/10.1103/PhysRevA.75.032110}{Evidence for the
  epistemic view of quantum states: A toy theory}, Phys. Rev. A 75 (2007)
  032110.
\newblock \href {http://dx.doi.org/10.1103/PhysRevA.75.032110}
  {\path{doi:10.1103/PhysRevA.75.032110}}.
\newline\urlprefix\url{http://link.aps.org/doi/10.1103/PhysRevA.75.032110}

\bibitem{chiribella2012quantum}
G.~Chiribella, G.~M. D?Ariano, P.~Perinotti, Quantum theory, namely the pure
  and reversible theory of information, Entropy 14~(10) (2012) 1877--1893.

\bibitem{coecke2013causal}
B.~Coecke, R.~Lal, Causal categories: relativistically interacting processes,
  Foundations of Physics 43~(4) (2013) 458--501.

\bibitem{coecke2014terminality}
B.~Coecke, Terminality implies non-signalling, arXiv preprint arXiv:1405.3681.

\bibitem{stueckelberg1960quantum}
E.~C. Stueckelberg, Quantum theory in real hilbert space, Helv. Phys. Acta
  33~(727) (1960) 458.

\bibitem{araki1980characterization}
H.~Araki, On a characterization of the state space of quantum mechanics,
  Communications in Mathematical Physics 75~(1) (1980) 1--24.

\bibitem{wootters1990local}
W.~K. Wootters, Local accessibility of quantum states, Complexity, entropy and
  the physics of information 8 (1990) 39--46.

\bibitem{hardy2012limited}
L.~Hardy, W.~K. Wootters, Limited holism and real-vector-space quantum theory,
  Foundations of Physics 42~(3) (2012) 454--473.

\bibitem{shannon}
C.~Shannon, A mathematical theory of communication, Bell System Technical
  Journal, The 27~(3) (1948) 379--423.
\newblock \href {http://dx.doi.org/10.1002/j.1538-7305.1948.tb01338.x}
  {\path{doi:10.1002/j.1538-7305.1948.tb01338.x}}.

\bibitem{schumacher1995quantum}
B.~Schumacher, Quantum coding, Physical Review A 51~(4) (1995) 2738.

\bibitem{bennett}
C.~H. Bennett, More about entanglement and cryptography,
  \url{http://www.lancaster.ac.uk/users/esqn/windsor07/Lectures/Bennett2.pdf},
  accessed: 2014-11-14 (2007).

\bibitem{GN}
I.~M. Gelfand, M.~A. Naimark, On the imbedding of normed rings into the ring of
  operators in hilbert space, Matematiceskij sbornik 54~(2) (1943) 197--217.

\bibitem{S}
I.~E. Segal, Irreducible representations of operator algebras, Bulletin of the
  American Mathematical Society 53~(2) (1947) 73--88.
\newblock \href {http://dx.doi.org/10.1090/S0002-9904-1947-08742-5}
  {\path{doi:10.1090/S0002-9904-1947-08742-5}}.

\bibitem{everett1957relative}
H.~Everett~III, ``relative state" formulation of quantum mechanics, Reviews of
  modern physics 29~(3) (1957) 454.

\bibitem{barnum2013ensemble}
H.~Barnum, C.~P. Gaebler, A.~Wilce, Ensemble steering, weak self-duality, and
  the structure of probabilistic theories, Foundations of Physics 43~(12)
  (2013) 1411--1427.

\bibitem{choi1975completely}
M.-D. Choi, Completely positive linear maps on complex matrices, Linear Algebra
  and its Applications 10~(3) (1975) 285--290.

\end{thebibliography}

\end{document}